\documentclass[twocolumn,preprintnumbers,amsmath,amssymb,nofootinbib]{revtex4-2}

\usepackage[utf8]{inputenc}
\usepackage[]{hyperref}
\usepackage[]{graphicx}
\usepackage{amsmath,amssymb,xfrac,multirow}
\usepackage{float}
\usepackage{xcolor}

\usepackage{miller}

\usepackage{scalerel}

\newcommand{\etal}{\textit{et al}.}
\newcommand{\ie}{\textit{i.e.}}
\newcommand{\eg}{\textit{e.g.}}

\newcommand{\magnetite}{Fe$_3$O$_4$}
\newcommand{\hematite}{Fe$_2$O$_3$}
\newcommand{\wustite}{Fe$_{1-x}$O}
\newcommand{\dftu}{$\text{DFT}+U$}

\begin{document}


\title{Development of an Atomic Cluster Expansion potential for iron and its oxides}

\author{Baptiste Bienvenu}%
 \email{b.bienvenu@mpie.de}
 \affiliation{Max Planck Institute for Sustainable Materials, Max-Planck-Straße 1, 40237 Düsseldorf.}%
\author{Mira Todorova}%
 \email{m.todorova@mpie.de}%
 \affiliation{Max Planck Institute for Sustainable Materials, Max-Planck-Straße 1, 40237 Düsseldorf.}
\author{Jörg Neugebauer}%
 \email{j.neugebauer@mpie.de}%
 \affiliation{Max Planck Institute for Sustainable Materials, Max-Planck-Straße 1, 40237 Düsseldorf.}
\author{Dierk Raabe}%
 \email{d.raabe@mpie.de}%
 \affiliation{Max Planck Institute for Sustainable Materials, Max-Planck-Straße 1, 40237 Düsseldorf.}
\author{Matous Mrovec}
 \affiliation{Interdisciplinary Centre for Advanced Materials Simulations, Ruhr Universität Bochum, 44780 Bochum.}
\author{Yury Lysogorskiy}
 \affiliation{Interdisciplinary Centre for Advanced Materials Simulations, Ruhr Universität Bochum, 44780 Bochum.}
\author{Ralf Drautz}
 \affiliation{Interdisciplinary Centre for Advanced Materials Simulations, Ruhr Universität Bochum, 44780 Bochum.}
 

\begin{abstract}

The combined structural and electronic complexity of iron oxides poses many challenges to atomistic modeling. To leverage limitations in terms of the accessible length and time scales, one requires a physically justified interatomic potential which is accurate to correctly account for the complexity of iron-oxygen systems. Such a potential is not yet available in the literature. In this work, we propose a machine-learning potential based on the Atomic Cluster Expansion for modeling the iron-oxygen system, which explicitly accounts for magnetism. We test the potential on a wide range of properties of iron and its oxides, and demonstrate its ability to describe the thermodynamics of systems spanning the whole range of oxygen content and including magnetic degrees of freedom.


\end{abstract}


\maketitle

\noindent
\textsc{\textbf{INTRODUCTION}}
\noindent

\noindent
Iron is one of the most abundant elements on Earth and is the primary constituent in steels and other metallic alloys. In both its natural and processed appearances, it is often present in the form of oxides, which develop easily under ambient conditions. A thorough understanding of properties of iron oxides down to the atomic and electronic levels is therefore crucial for optimizing the production of pure iron from its ores as well as for minimizing the impact of environmental degradation due to oxidation.

The Fe-O binary system, although it contains only a few stable compounds, reveals a wide range of structural and electronic variety and complexity \cite{Cornell2003}. Iron oxides are mainly found in nature in the form of three stable minerals, namely, wüstite (FeO), magnetite ({\magnetite}) and hematite ({\hematite}), by increasing oxygen concentration.

Most atomistic studies of the Fe-O system were carried out using Density Functional Theory (DFT) calculations. However, a faithful description of the respective compounds requires the use of different exchange and correlation (xc-) functionals depending on the oxygen content in the structure of interest. Indeed, standard semi-local DFT functionals such as LDA or various GGAs predict not only Fe, but also its oxides to be metallic, which is in disagreement with experimental observations. This discrepancy has motivated further investigations of the oxides with more sophisticated methodologies, such as {\dftu} or hybrid functionals \cite{Meng2016}.

First-principles calculations face limitations regarding the size of simulated systems and the time scales of simulations. Therefore, they are impractical for exhaustive studies of extended defects, such as grain boundaries, interfaces, dislocations or kinetic processes, e.g., diffusion, phase transformation and oxidation. Atomistic modeling of such complex phenomena necessitates an interatomic potential that is both accurate and efficient. However, the development of robust and transferable potentials that would be able to capture the Fe-O system in its entirety over a broad range of conditions of temperature, stoichiometry, stress, chemical potential, etc., presents a major challenge.

Most of the available interatomic potentials for the Fe-O system are based on the ReaxFF formalism \cite{Chenoweth2008} and show severe discrepancies with DFT and experiments in terms of structural and transport properties \cite{Thijs2023}. Most importantly, the parameterization of Aryanpoor {\etal} \cite{Aryanpour2010} and subsequent models predict the three iron oxides to be dynamically unstable. An analytical bond order potential (ABOP) was developed recently  \cite{Byggmastar2019} to investigate oxygen defects in BCC Fe and the stoichiometric wüstite phase. However, due to its limited transferability it yields non-physical results for phases with higher oxygen content, with {\magnetite} and {\hematite} melting already at room temperature \cite{Byggmastar2019}.


In the present work, we develop a machine-learned interatomic potential (MLIP) based on the Atomic Cluster Expansion (ACE) \cite{Drautz2019} that explicitly accounts for magnetism, which is crucial for a physically correct description of the Fe-O system. The potential is fitted to an extensive DFT database encompassing pure Fe as well as a large variety of oxygen containing phases. We validate the ACE parametrization for a wide range of properties of both pristine and defective phases spanning the whole range of oxygen concentrations.  We also demonstrate the ability of the model to provide an explicit description of magnetic degrees of freedom and behavior at finite temperatures.\\

\noindent
\textsc{\textbf{RESULTS}}

\noindent
\textbf{DFT reference data}

\noindent
\textit{The Fe-O system}

\begin{figure}[!htb]
    \includegraphics[trim = 0mm 1mm 0mm 0mm, clip, width=0.95\linewidth]{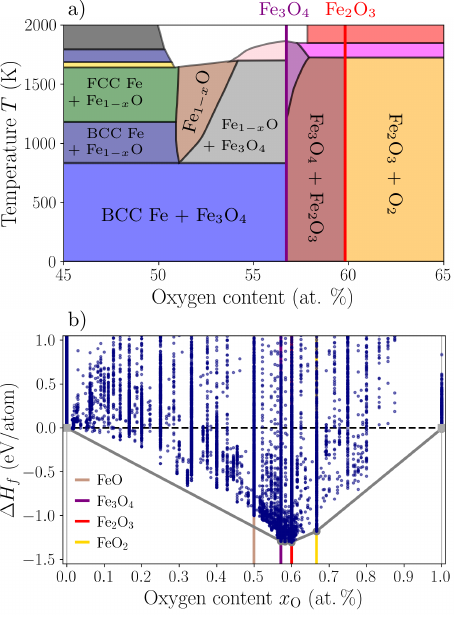}
    \caption{a) Experimental Fe-O phase diagram adapted from Ref. \cite{Darken1946} showing the stable compounds. b) DFT convex Hull ({\ie} formation enthalpy $\Delta H_f$ as a function of the oxygen atomic concentration $x_{\text{O}}$) computed in this work.}
    \label{fig:ref}
\end{figure}

\noindent
Target systems for the potential span the experimental Fe-O binary phase diagram presented in Fig. \ref{fig:ref}\,a \cite{Darken1946}, from pure Fe to highly oxidized environments, encompassing the stability ranges of the three iron oxides.


At low pressure and temperature, pure Fe crystallizes in a body-centered cubic (BCC) phase with ferromagnetic (FM) order. This structure is stable up to approximately 1200\,K, where the BCC FM phase destabilizes in favor of a disordered paramagnetic (PM) face-centered cubic (FCC) phase. At higher pressures, pure Fe takes a hexagonal close-packed (HCP) structure, which remains stable up to extremely high pressures \cite{Dorogokupets2017}. 

In the binary Fe-O system, up to its solubility limit, O atoms occupy interstitial sites in the host Fe lattice. The first ordered oxide above this limit, wüstite {\wustite}, appears in the phase diagram at approximately $50\%$ of oxygen and is characterised by a degree of Fe deficiency $x$ varying between 17 and 5$\,\%$ within its stability range \cite{Hazen1984}. {\wustite} has a rock-salt structure (NaCl, Fm$\overline{3}$m), with two Fe and O intertwined FCC sub-lattices and an antiferromagnetic (AF) order at low temperature. At $57\%$ of oxygen, the stable oxide becomes magnetite, {\magnetite}, which crystallizes under ambient conditions in an inverse spinel structure (Fd$\overline{3}$m) with a ferrimagnetic order. In {\magnetite}, O atoms are arranged on a FCC lattice, while the Fe atoms sit in both octahedral and tetrahedral sites, with the two sites having opposite spin directions. Above $60\%$ of oxygen, {\hematite} is the primary stable oxide, with a corundum structure (R$\overline{3}$c) with AF magnetic order and a HCP underlying O sub-lattice \cite{Cornell2003}. 

In terms of electronic properties, under ambient temperature and pressure, {\wustite} and {\hematite} are insulators, while {\magnetite} is a half-metal \cite{Cornell2003}, adding a further degree of complexity to the structural and magnetic complexity highlighted above.\\

\noindent
\textit{DFT calculations}

\noindent
Carrying out atomistic simulations of iron oxides at the DFT level is challenging due to the high degree of electronic correlations of the $d$-orbitals on Fe atoms. A proper description of the electronic properties thus needs to rely on a formalism that is able to incorporate electronic correlations to some extent, for instance, via the widely used framework of {\dftu} \cite{Dudarev1998}. However, a better agreement with experiments in terms of electronic properties for iron oxides \cite{Meng2016} is counteracted by a worse description of structural and thermodynamic properties; for instance, predicting an almost zero diffusion barrier for oxygen interstitials in BCC Fe \cite{Shang2014} or incorrect ordering of the BCC and FCC Fe phases (see Supplementary Materials). This is partly due to the fact that the $U$ correction is an empirical parameter, which is usually tuned to reproduce a given property of interest ({\eg} the electronic band gap), and is thus material-dependent \cite{Meng2016}. A common compromise to obtain properties of both pure Fe and its oxides within DFT is to use {\dftu} for oxides and a standard functional, such as GGA-PBE \cite{Perdew1996}, for pure Fe. An empirical mixing scheme is then applied to obtain, for instance, formation enthalpies comparable to experiments \cite{Jain2011}, which is implemented in the MaterialsProject database \cite{Jain2013}.

However, in the context of fitting a MLIP, a key feature of the underlying DFT reference data is its coherency, mainly in terms of DFT parameters, and in the present case of the Fe-O system, most particularly in terms of xc-functional. Thus, to have a consistent description of Fe atoms across the whole range of oxygen concentration, which is needed to fit the potential but also motivated by the physical intuition that the mixing scheme discussed above is not applicable for structures found in-between pure Fe and its oxides, a trade-off is needed and a single DFT functional must be chosen. Given that GGA-PBE gives satisfactory agreement with experiments in terms of structural properties of the three oxides (lattice constants and elastic properties, see Tab. \ref{tab:bulk}) while offering a good description of metallic Fe, we chose to use this xc-functional to compute the whole DFT dataset.

Since Fe and its oxides exhibit ordered magnetic states from 0\,K up to relatively high temperatures, magnetism must be accounted for in all DFT calculations. The Curie temperature of Fe is around 1000\,K where the BCC FM order transitions to a disordered PM state. For both {\magnetite} and {\hematite}, the Néel temperature marking the transition to the PM state is about 950\,K \cite{Samara1969,Neskovic1977}. As for wüstite ({\wustite}), the Néel temperature is approximately 200\,K \cite{Cornell2003}, which depends on the degree of Fe depletion $x$ \cite{McCammon1992}. Thus, under normal conditions for various applications such as catalysis (around room temperature \cite{Pereira2012}) and oxidation/reduction (up to 1000\,K), both iron and its oxides still have an ordered magnetic state, motivating  the explicit consideration of magnetism in the present potential. To this extent, magnetism is included in its collinear state in all DFT calculations. More details about the DFT parameters are given in Methods.\\


\noindent
\textit{Reference database}
\label{sec:database}

\noindent
From the experimental phase diagram of the binary Fe-O system presented in Fig. \ref{fig:ref}.a, we identify the most important prototype structures as the stable compounds, which are pure Fe, wüstite ({\wustite}), {\magnetite} and {\hematite}. These materials constitute the foundation for the DFT reference data. Pure oxygen is included through sampling of its gas and liquid phases only and in much lesser detail compared to structures containing up to 80$\%$ of oxygen. This is also motivated by the poor description of molecular oxygen binding by the GGA-PBE xc-functional within DFT.

All prototype structures are then distorted by applying random strains to the unit cells to describe their elastic properties and deformation behavior. Additionally, atoms are randomly displaced to account for thermal vibrations. After this first layer for the construction of the database, the most relevant defects are included in these materials, mainly vacancies, interstitials (both Fe and O), clusters composed of the two types of point defects and surfaces of various low-index orientations. Additionally, some grain boundary and dislocation configurations are included for pure Fe in its ground-state BCC structure. 

Finally, to sample a wider part of the configuration space, both in terms of atomic environments and oxygen concentrations, we also included randomly generated structures in the training data. These structures are generated such that they respect constraints on their point-group symmetry, number of atoms, and oxygen concentration \cite{Pickard2011}. These "random" structures are of great importance for the parametrization of the present ACE potential, since they allow the model to confidently describe out-of-equilibrium and highly distorted structures found for instance at the core of complex defects, during transitions between different phases, or in the liquid phases. It has been shown that including such structures in the reference database drastically increases the reliability and transferability of MLIPs \cite{Poul2023}. Since the aim of the present ACE potential is to describe structures which are for the most part not known \textit{a priori} and span the whole range of oxygen concentration, it is important that the model can interpolate between data points for a wide variety of atomic environments. These structures make for the majority of the overall database.

Additionally, we also sampled the magnetic degrees of freedom by considering different collinear magnetic orders for both prototype materials ({\ie} Fe and its oxides) and during the generation of random structures. The final DFT reference database is presented in Fig. \ref{fig:ref}\,b, showing the formation enthalpy $\Delta H_f$ of all structures as a function of the oxygen concentration $x_{\text{O}}$.

We note that stoichiometric wüstite (FeO) is not part of the DFT convex hull, thus predicted thermodynamically unstable at 0\,K, in agreement with experiments, but not with {\dftu} calculations \cite{Meng2016}. This observation is however difficult to rationalize through comparison with experiments, since the cation defect structure in {\wustite} is extremely complicated and still under debate \cite{Hazen1984}. The lowest formation enthalpy is found for {\hematite}, in agreement with experiments \cite{Stull1971}. We also report the stability of FeO$_2$, a compound observed experimentally at high temperature and pressure \cite{Hu2016}.\\

\noindent
\textbf{Potential parameterization}

\noindent
In this work, we propose a MLIP for the Fe-O system based on the Atomic Cluster Expansion (ACE) \cite{Drautz2019}, which provides a physically-justified and complete basis set for the description of atomic environments. Potentials based on the ACE have proven to be robust and versatile to capture different types of interactions, such as highly covalent carbon \cite{Qamar2023}. They are also able to incorporate additional degrees of freedom like magnetic moments, as seen, {\eg}, in a model which was parameterized for elemental Fe including non-collinear magnetism \cite{Rinaldi2024}. For a detailed description of the ACE formalism, please refer to Refs. \cite{Drautz2019, Bochkarev2022, Lysogorskiy2021}.\\

\noindent
\textit{Explicit account of magnetism}
\label{sec:ace_mag}

\noindent
As discussed above, most materials of interest in the Fe-O phase diagram show ordered magnetic states up to temperatures of around 1000\,K \cite{Cornell2003}. An explicit account of magnetism in the model therefore appears necessary to reliably describe the system at various temperatures.

In this work, we propose to include the magnetic degrees of freedom in the ACE model in a rather simple way by using an Ising-like description. To do so, we fit the present potential considering three different types of Fe atoms defined according to the sign of their magnetic moments: spin up ($\text{Fe}_{\uparrow}$), spin down ($\text{Fe}_{\downarrow}$) and non-magnetic ($\text{Fe}_{\text{NM}}$). A threshold absolute value of $0.1\,\mu_{\text{B}}$ is set to distinguish between these three types of Fe atoms. Such a simple model for magnetism does not provide an accurate representation of complex magnetic excitations at finite temperature, which are mostly non-collinear. However, it allows for an  explicit account of magnetism at a cost and complexity compatible with large-scale simulations of magnetic systems where both atomic positions and magnetic moments can evolve. A more accurate account of magnetic degrees of freedom within the ACE formalism was recently proposed for pure Fe \cite{Rinaldi2024}, showing excellent description of magnetic excitations, allowing for instance to accurately predict the transition temperatures from BCC to FCC, which is driven by magnetic disorder. However, the training data for such models require the use of constrained magnetic DFT calculations and a thorough sampling of magnetic degrees of freedom, which is not compatible in terms of number of configurations with the scope of the current potential, which is aimed at describing a wide variety of atomic environments. It also yields a rather fast potential compared to the full account of magnetism, allowing for large-scale simulations at a reduced cost.

Magnetic structures are included twice in the training set, considering in each case opposite spin configurations (for instance, FM order with all spins up or all spins down) to enforce the inversion symmetry during the fitting procedure detailed in the next section. The "doubled" dataset contains approximately 40000 structures.\\

\noindent
\textit{Model parameters and accuracy}

\noindent
A cutoff distance common to all interactions was set to 7.0\,{\AA}, which we find is necessary to capture the range of the different interactions between Fe and O atoms in various bonding environments. The present model contains 1000 functions per element, which, including all three types of iron atoms ($\text{Fe}_{\uparrow}$, $\text{Fe}_{\downarrow}$ and $\text{Fe}_{\text{NM}}$) and oxygen, results in a total of 4000 functions and 10352 parameters. The optimization of the parameters was done using the \textsc{Pacemaker} code \cite{Lysogorskiy2021,Bochkarev2022}.

The overall accuracy of the potential across the entire database (in mean absolute errors) is 28\,meV/atom and 78\,meV/{\AA} in energies and force components respectively. During the fitting procedure, structures lying within 1\,eV/atom of the Fe-O DFT convex Hull (see Fig. \ref{fig:ref}\,b) were given greater weights, resulting in an improved accuracy of 18\,meV/atom and 61\,meV/{\AA} in energies and force components, respectively, in this region. These seemingly large errors mostly originate from trying to fit complex magnetic excitations with a simple model for magnetism. We will show in the following that however large the magnitude of the reported error metrics seems, it does not reflect the accuracy of the ACE potential in terms of the predicted properties.\\

\noindent
\textbf{Validation on bulk properties}

\noindent
In this section, we validate the accuracy of the present ACE potential on predicting bulk properties of both Fe and its oxides, by comparing the results to both DFT data and experiments. We also compare the results obtained with the ACE potential to the predictions of the two Fe-O interatomic potentials available from the literature, namely the ABOP potential from Byggmastar {\etal} \cite{Byggmastar2019} referred to as ABOP2019, and the reaxFF potential of Aryanpoor {\etal} \cite{Aryanpour2010} referred to as reaxFF2010. 

\begin{center}
    \begin{table}[!htb]
        \caption{Bulk properties of Fe and its oxides predicted by ACE and compared to DFT calculations (with and without $U$ correction) and experiments (extrapolated to 0\,K): lattice constants ($a_0$ for cubic lattices, $a$ and $c$ for hexagonal {\hematite}), bulk modulus $B_0$, and 0\,K formation enthalpy $\Delta H_f$. For each structure, the lowest energy crystal structure and magnetic order are indicated (FM for ferromagnetic, AF for antiferromagnetic and FeM for ferrimagnetic). The {\dftu} values for $\Delta H_f$ are obtained without applying the correction schemes of Refs. \cite{Jain2011,Wang2021}. Experimental data are taken from various references, indicated for each property and material.}
        \label{tab:bulk}
        \centering
        \begin{tabular}{l c c c l}
              & & \multicolumn{2}{c}{DFT} & \\
              & ACE \, & \multicolumn{1}{c}{GGA-PBE} \, & \multicolumn{1}{c}{$+U_{\text{Fe}}=4$\,eV} \, & Expt. \\
            \hline
            \multicolumn{5}{l}{Iron, Fe (BCC, FM)} \\
            \hline
            $a_0$ ({\AA}) & 2.83 & 2.83 & 2.95 & 2.86 \cite{Basinski1955} \\
            $B_0$ (GPa) & 163 & 188 & 123 & 164 \cite{Dorogokupets2017} \\
            \hline
            \multicolumn{5}{l}{Stoichiometric wüstite, FeO (distorted NaCl, AF)} \\
            \hline
            $a_0$ ({\AA}) & 4.27 & 4.26 & 4.31 & 4.33 \cite{Zhang2000} \\
            $B_0$ (GPa) & 183 & 195 & 164 & 175 \cite{Zhang2000} \\
            $\Delta H_f$ (eV/atom) & $-0.93$ & $-0.91$ & $-1.43$ & $-1.41$ \cite{Stull1971} \\
            \hline
            \multicolumn{5}{l}{Magnetite, {\magnetite} (inverse spinel, FeM)} \\
            \hline
            $a_0$ ({\AA}) & 8.39 & 8.39 & 8.47 & 8.40 \cite{Haavik2000} \\
            $B_0$ (GPa) & 197 & 172 & 191 & 183 \cite{Haavik2000} \\
            $\Delta H_f$ (eV/atom) & $-1.29$ & $-1.29$ & $-1.75$ & $-1.66$ \cite{Stull1971} \\
            \hline
            \multicolumn{5}{l}{Hematite, {\hematite} (corundum, AF)} \\
            \hline
            $a$ ({\AA}) & 5.00 & 5.02 & 5.07 & 5.04 \cite{Finger1980} \\
            $c$ ({\AA}) & 13.94 & 13.90 & 13.90 & 13.75 \cite{Finger1980} \\
            $B_0$ (GPa) & 165 & 172 & 191 & 225 \cite{Finger1980} \\
            $\Delta H_f$ (eV/atom) & $-1.30$ & $-1.30$ & $-1.81$ & $-1.71$ \cite{Stull1971} \\
            \hline
        \end{tabular}
    \end{table}
\end{center}

A summary of the bulk properties of pure Fe and its three oxides (lattice constants, bulk moduli and formation enthalpies) is presented in Tab. \ref{tab:bulk}. A very good agreement between the properties predicted by ACE and both DFT and experimental references is obtained in terms of lattice constants and bulk moduli for all four materials. The main discrepancy between ACE and experiments, which is also reflected in the underlying DFT data used for its fitting, is the underestimation of the 0\,K formation enthalpies of the iron oxides. This discrepancy is partially resolved by using {\dftu} as discussed above. Hereby, a value of $U_{\text{Fe}}=4$\,eV was chosen for comparison since it has been shown to give the most satisfactory agreement with experiments for all three oxides in terms of electronic properties and formation energies \cite{Meng2016}. Otherwise, we also note that for structural properties, GGA-PBE, and thus the ACE potential, are in better agreement with experiments than predictions of {\dftu}.


\begin{figure*}[!htb]
    \hspace{0mm}
    \includegraphics[trim = 0mm 0mm 0mm 0mm, clip, width=1\linewidth]{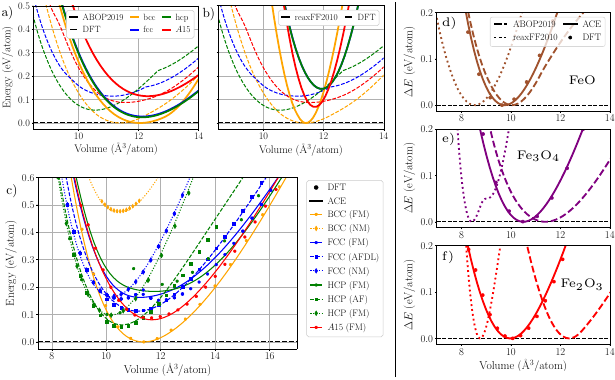}
    \caption{(a-c) Energy-volume curves for different crystal structures for pure Fe, namely BCC, FCC, HCP and $A15$, comparing DFT (dashed lines) with a) ABOP2019, b) reaxFF2010 and c) the present Fe-O ACE potential. For the DFT data, only the magnetic ground-state is plotted at each volume in a) and b) since the two potentials cannot distinguish between different magnetic orders. This reflects in the kinks present in the DFT curves for FCC and HCP structures, corresponding to a change in the lowest energy magnetic order. In c), different magnetic orders (AFDL: antiferromagnetic double layer; NM: non-magnetic) are shown when comparing DFT to ACE since the potential captures magnetic degrees of freedom. (d-f) Energy-volume curves for the three iron oxides: d) stoichiometric wüstite FeO, e) magnetite {\magnetite}, and f) hematite {\hematite} obtained with the ABOP2019 (dashed lines), reaxFF2010 (dotted lines), ACE (full lines) potentials, and DFT (GGA-PBE, symbols).}
    \label{fig:bulk}
\end{figure*}

We present in Fig. \ref{fig:bulk} the variation of the energy as a function of the atomic volume for different crystal structures and magnetic orders of Fe and the three FeO, {\magnetite} and {\hematite} oxides. We report a very good agreement between the DFT reference and the predictions of the ACE potential, capturing the hierarchy between both different crystal structures and magnetic orders, as exemplified in the case of pure Fe (see Fig. \ref{fig:bulk}\,c). This complex intertwining of phases is not well described by the two ABOP2019 and reaxFF2010 potentials, as presented in Figs. \ref{fig:bulk}.a and b respectively, which are not able to distinguish the FCC from the HCP structures and also predict the wrong ordering of phases. The ACE predictions also agree very well with the DFT reference for the three oxides, showing its applicability to both metallic bonding in pure Fe and covalent bonding in the oxides. 

For the oxides, reaxFF2010 also performs poorly, showing two metastable equilibrium volumes for {\magnetite} and predicting bulk moduli much higher than the DFT reference, except for FeO. ABOP2019 predicts good properties for FeO, which was included in its training set, while predictions for {\magnetite} and {\hematite} are less reliable since the potential was not trained to reproduce their properties. Most particularly, the authors reported these two oxides to melt at room temperature. As for reaxFF2010, we note that all three iron oxides (FeO, {\magnetite} and {\hematite}) are found to be dynamically unstable (see Supplementary Materials), showing imaginary phonon modes.

For the above-mentioned reasons, further comparisons including the two reaxFF2010 and ABOP2019 potentials will be omitted when discussing defect properties in iron and its oxides in the following sections.

We also computed phonon spectra for the stable polymorphs of pure Fe, as well as for {\magnetite} and {\hematite}, and compared the results obtained with the ACE potential to DFT calculations. The results are presented in Supplementary Materials (see Figs. 2 and 6) and again show a very satisfactory description of these vibrational properties by the present ACE potential.\\\\

\noindent
\textbf{Point defects and diffusion}


\noindent
\textit{Point defect formation energies}

\begin{figure}[!htb]
    \hspace{0mm}
    \includegraphics[trim = 0mm 0mm 0mm 0mm, clip, width=1\linewidth]{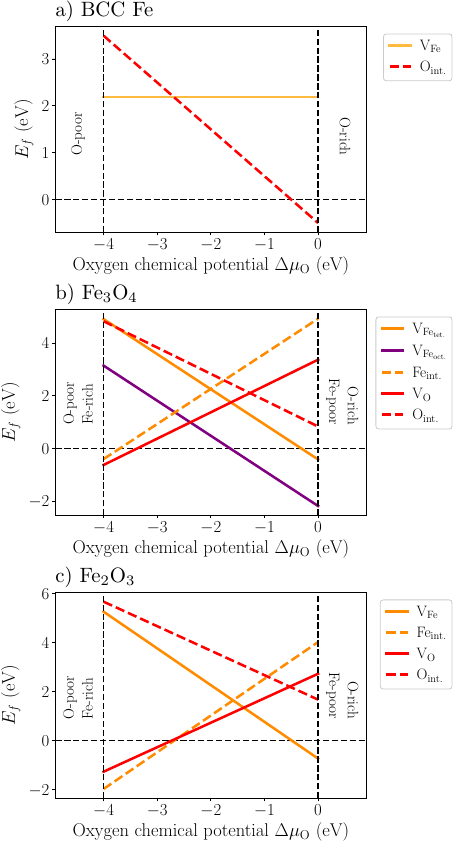}
    \caption{Point defect formation energies $E_f$ as a function of the oxygen chemical potential change $\Delta \mu_{\text{O}}$ in a) BCC Fe (O octahedral interstitials only), b) {\magnetite}, and c) {\hematite} predicted by ACE. $\Delta \mu_{\text{O}}$ is referenced to the O$_2$ molecule.}
    \label{fig:muO_defects}
\end{figure}

\noindent
The formation energies of vacancies and interstitials as a function of the oxygen chemical potential $\Delta \mu_{\text{O}}$ change (referenced to the energy of an O atom in molecular O$_2$) is presented in Fig. \ref{fig:muO_defects} for BCC Fe, {\magnetite} and {\hematite}. Wüstite, and in particular its stoichiometric form FeO, is not included since it is unstable at 0\,K according to both DFT calculations and the ACE potential. Details about the evaluation of the formation energies are given in Methods. Only results obtained with ACE are shown. Also, it is worth noting that in both DFT and the present ACE potential, the effect of charges, which have been shown to influence the formation energies of point defects in oxides with respect to the neutral case, most particular for Fe vacancies and interstitials \cite{Banerjee2023}, is not included.


\noindent
\textit{BCC iron}

\begin{figure}[!htb]
    \hspace{-10mm}
    \includegraphics[trim = 0mm 0mm 0mm 0mm, clip, width=0.85\linewidth]{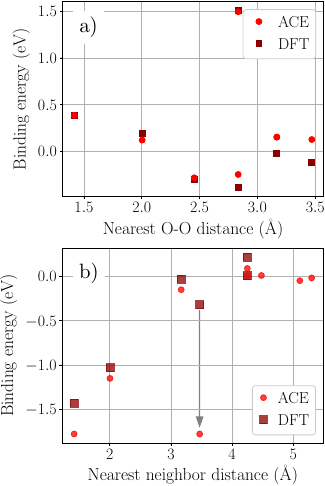}
    \caption{Binding energies between a) two octahedral O interstitials, and b) one O octahedral interstitial and an Fe vacancy as a function of the distance between the two point defects. DFT data in b) are taken from Ref. \cite{Barouh2014}. The grey arrow at the position of 4$^{\text{th}}$ nearest neighbor indicates its relaxation to a 1$^{\text{st}}$ nearest neighbor configuration.}
    \label{fig:binding_bccFe_O}
\end{figure}

\noindent
The formation energy of a single vacancy in BCC Fe is 2.2\,eV according to ACE, showing a perfect agreement with the DFT reference of 2.2\,eV and the experimental value of $2.0\pm 0.2$\,eV \cite{Schepper1983}. The formation energy of an oxygen interstitial in BCC Fe, which prefers to sit on the octahedral sites of the lattice as previously reported \cite{Barouh2014}, is also in very good agreement with the reference DFT and the literature. We also present on Fig. \ref{fig:binding_bccFe_O} the binding energy between two O interstitial atoms in BCC Fe, as well as the binding energy between one O interstitial atom and a Fe vacancy. These properties play a key role in the diffusion of both oxygen in iron \cite{Shang2014,Barouh2014} and in the formation of oxides in the oxygen-rich region \cite{Fu2007}.


We report a very satisfactory agreement between ACE and available DFT reference for both O-O and O-vacancy binding energies, reproducing the strongly attractive and short-range interaction between oxygen atoms and vacancies in BCC Fe, as well as the repulsive interaction between neighboring oxygen interstitials. For the 4th nearest neighbor site between an octahedral O and an Fe vacancy (distance of approximately 3.5\,{\AA} in Fig. \ref{fig:binding_bccFe_O}\,b), ACE predicts the configuration to be unstable and relaxes to a 1st nearest neighbor position. The latter was also reported in another DFT study \cite{Wang2020}.

We also show in Tab. \ref{tab:diff_barriers_bccFe} the 0\,K migration barriers of Fe vacancies and O interstitials for different pathways, and compare the results obtained with ACE with available DFT data. The migration energy of a single Fe vacancy is 0.73\,eV and in very good agreement with previous DFT studies \cite{Fu2005,Rinaldi2024}.


\begin{figure*}[!htb]    
    \hspace{-12mm}
    \includegraphics[trim = 0mm 0mm 0mm 0mm, clip, width=0.75\linewidth]{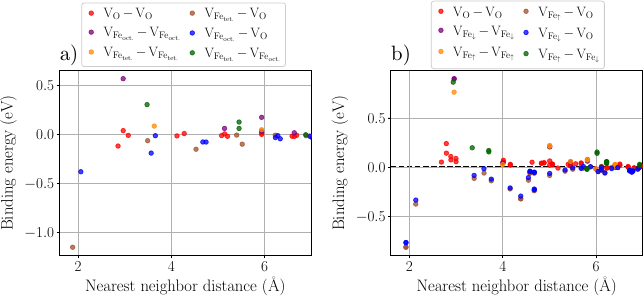}
    \caption{Binding energies between two vacancies in a) {\magnetite} for all combinations of O, tetrahedral and octahedral Fe vacancies, and b) {\hematite} for all combinations of O, spin up and spin down Fe vacancies, both obtained using the ACE potential.}
    \label{fig:binding_oxides}
\end{figure*}

The diffusion barrier of an O interstitial atom between two adjacent octahedral sites, going through the unstable tetrahedral position, is predicted by the ACE potential at 0.44\,eV, which compares very well with the DFT reference value of 0.46\,eV computed in this work, but also with the one reported in previous DFT studies \cite{Barouh2014,Shang2014,Wang2020} (see Tab. \ref{tab:diff_barriers_bccFe}). We also compare the diffusion barrier for a "cage-jump", {\ie} O migration between two octahedral sites which are nearest neighbors to a Fe vacancy. The height of the barrier obtained with ACE is 0.63\,eV, which compares very well with the 0.59\,eV value reported in a previous DFT study \cite{Wang2020}.

\begin{center}
    \begin{table}[!htb]
        \caption{Height of diffusion barriers (in eV) for migration of O interstitial atoms and Fe vacancy in BCC Fe (FM order): O$_{\text{oct.}}\,-\,$O$_{\text{oct.}}$ corresponds to the jump of an O interstitial atom between two adjacent octahedral sites; O$_{\text{oct.}}^{\text{V, 1NN}}\,-\,$O$_{\text{oct.}}^{\text{V, 1NN}}$ corresponds to the same jump but between two adjacent octahedral sites which are nearest neighbors to a Fe vacancy; 1V$_{\text{Fe}}^{\text{1NN}}\,-\,$1V$_{\text{Fe}}^{\text{1NN}}$ corresponds to the migration of a single Fe vacancy; 2V$_{\text{Fe}}^{\text{1NN}}\,-\,$2V$_{\text{Fe}}^{\text{1NN}}$ is the collective migration of a di-vacancy, and 3V$_{\text{Fe}}^{\text{1NN}}\,-\,$3V$_{\text{Fe}}^{\text{1NN}}$ of a tri-vacancy.}
        \label{tab:diff_barriers_bccFe}
        \centering
        \begin{tabular}{l c l}
              & ACE \, & DFT \\
            \hline
            O$_{\text{oct.}}\,-\,$O$_{\text{oct.}}$ & 0.44 & 0.46, 0.56 \cite{Barouh2015}, 0.53 \cite{Shang2014}, 0.51 \cite{Wang2020} \\
            O$_{\text{oct.}}^{\text{V, 1NN}}\,-\,$O$_{\text{oct.}}^{\text{V, 1NN}}$ & 0.63 & 0.59 \cite{Wang2020} \\
            \hline
            1V$_{\text{Fe}}^{\text{1NN}}\,-\,$1V$_{\text{Fe}}^{\text{1NN}}$ & 0.70 & 0.67 \cite{Fu2005,Rinaldi2024} \\
            2V$_{\text{Fe}}^{\text{1NN}}\,-\,$2V$_{\text{Fe}}^{\text{1NN}}$ & 0.69 & 0.62 \cite{Fu2005} \\
            3V$_{\text{Fe}}^{\text{1NN}}\,-\,$3V$_{\text{Fe}}^{\text{1NN}}$ & 0.29 & 0.35 \cite{Fu2005} \\
            \hline
        \end{tabular}
    \end{table}
\end{center}

In Ref. \cite{Shang2014}, the authors state that including O migration through interstitial diffusion without Fe vacancies cannot account for the rather slow diffusion reported experimentally for BCC Fe. As reported previously \cite{Barouh2014} and also reproduced by the present ACE potential, the proximity to a Fe vacancy indeed raises the energy barrier for O interstitial diffusion. Since the interaction between an Fe vacancy and an O interstitial atom is strongly attractive (see Fig. \ref{fig:binding_bccFe_O}\,b), it is indeed likely that such coupled motion of the two defects influences the diffusion properties of O in BCC Fe, depending on the concentration of both defects for given environmental conditions (temperature and oxygen activity or pressure).

Extended validation on other defect properties in different polymorphs of pure Fe is also presented in Supplementary Materials, including surface energies, self-interstitial configurations, and vacancy formation energy in BCC, FCC and HCP Fe, considering different magnetic orders. We also assessed the ability of the potential to describe complex defects such as screw dislocations and grain boundaries in BCC Fe. We report a very satisfying agreement with reference DFT data, even for properties and structures that have not been explicitly included in the training set.\\



\noindent
\textit{Iron oxides}

\noindent
We present in Fig. \ref{fig:muO_defects}\,b and c the formation energy of vacancies and interstitials in both {\magnetite} and {\hematite} as a function of the oxygen chemical potential. As can be seen in the experimental phase diagram of Fig. \ref{fig:ref}\,a, {\magnetite} is slightly off-stoichiometric at high temperatures, with a cation Fe deficiency \cite{Dieckmann1983_V}. Among the two Fe lattice sites in {\magnetite}, the vacancy formation energy is lower for octahedral positions than for tetrahedral, which was also reported experimentally \cite{Dieckmann1977_I}, with ACE predicting an energy difference of 1.63\,eV between the two sites.

In {\magnetite}, there are three inequivalent interstitial sites: two tetrahedral and one octahedral with respect to the oxygen FCC sub-lattice \cite{Dieckmann1982_IV}. For each site, the formation energy was evaluated for an iron interstitial atom of both spin up and down configuration, as well as for an oxygen atom. Only the most favorable sites are shown on Fig. \ref{fig:muO_defects}. According to the ACE potential, the open tetrahedral site is the most stable for Fe interstitials, followed by the octahedral site with an energy difference of 0.96\,eV. The energy difference between the two spin states is very small, but a spin opposite to the surrounding Fe atoms is more stable. For O interstitials, the tetrahedral site is also the most stable according to ACE, the octahedral site having an energy higher by 0.09\,eV.

For {\hematite}, we report a higher formation energy of a Fe vacancy than for {\magnetite}. In {\hematite}, the most favourable site for both Fe and O atoms has an octahedral position with respect to the underlying HCP O sub-lattice.

We also present on Fig. \ref{fig:binding_oxides} the binding energies of different di-vacancies, for pairs of V$_{\text{Fe}}$-V$_{\text{Fe}}$, V$_{\text{Fe}}$-V$_{\text{O}}$ and V$_{\text{O}}$-V$_{\text{O}}$ in both {\magnetite} and {\hematite}. In both oxides, only pairs composed of one Fe and one O vacancy have a negative binding energy and only at short distances, indicating their tendency to bind and subsequently form clusters. 

\begin{figure}[!htb]    
    \includegraphics[trim = 0mm 0mm 0mm 0mm, clip, width=1\linewidth]{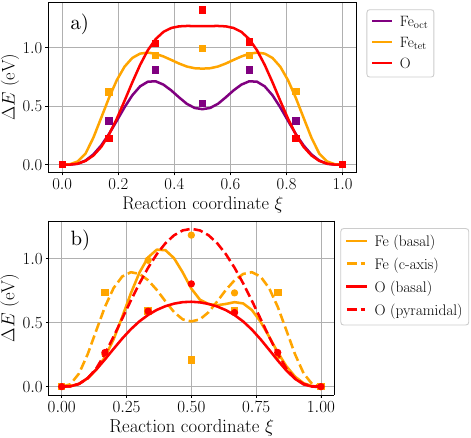}
    \caption{Migration barriers of Fe and O vacancies in a) {\magnetite}, and b) {$\alpha$-\hematite}. For {\magnetite}, both tetrahedral and octahedral Fe vacancies are considered, both moving in a \hkl{100} plane. For {\hematite}, Fe and O atoms moving in different planes of the hexagonal lattice are considered: in the basal plane for both Fe and O (solid lines), in the pyramidal plane for O (dashed line), and along the $c=\hkl[0001]$ axis for Fe (dashed line). All symbols correspond to DFT data computed in this work.}
    \label{fig:diff_ox}
\end{figure}

We next present vacancy migration barriers for the two oxides in Fig. \ref{fig:diff_ox}. Interstitial motion is more complex, and might involve spin flips as the cation moves through the material, which was reported in {\hematite} in a recent DFT study \cite{Banerjee2021}. Therefore, migration barriers are presented only for vacancies in the two {\magnetite} and {\hematite} oxides.


In {\magnetite}, the asymmetry between the two types of Fe vacancy (tetrahedral and octahedral) reported for the formation energy is also observed in the diffusion barriers, with the energy barrier for octahedral Fe vacancies being lower. Both results indicate that the vacancy-mediated diffusion of Fe in {\magnetite} is mainly due to the motion of octahedral vacancies, in line with experimental observations \cite{Dieckmann1977_I}.

For {\hematite}, Fe and O vacancies can diffuse along different paths: in the basal planes, the pyramidal planes, and along the $c=\hkl[0001]$ axis of the underlying hexagonal O sub-lattice. For Fe vacancies, ACE predicts an easier motion along the $c$ axis, with a slightly higher energy barrier in the basal planes. The migration barriers obtained with ACE agrees well with the DFT reference, except for the position located half-way along the path, which shows a lower energy in DFT. This is mainly due to a large reduction of the magnetic moment of the diffusing cation, which cannot be accounted for by the present ACE model. For O vacancies, ACE predicts an easier motion in the basal than in the pyramidal plane. The predicted barriers and their hierarchy is also in good agreement with recent {\dftu} results \cite{Banerjee2023}, where different charge states were considered for each defect.


From the presented diffusion barriers, one can see that oxygen vacancies can move more easily in {\hematite} than in {\magnetite}. Additionally, Fe vacancies have a lower migration barrier than O in {\magnetite}, while the barriers for the two species have similar heights in {\hematite}. While this can be linked to the transport properties of both species in the two oxides, it does not represent the full picture of diffusion, which must also account for the equilibrium concentration of defects.

For instance, assuming the vacancy-mediated diffusion of both Fe and O in {\hematite} to be solely described by the migration paths presented in Fig. \ref{fig:diff_ox}\,b, Fe and O appear to diffuse at a comparable rate, due to the similar height of the diffusion barrier. However, this is without considering the equilibrium concentration of defects, which, as can be seen in Fig. \ref{fig:muO_defects}\,c, strongly depends on the oxygen chemical potential. For instance, the formation energy of an O vacancy in the O-rich region is way higher than for Fe vacancies. Thus, in these conditions, the diffusivity of O in {\magnetite} will appear slower than Fe due to the comparatively low concentration of the defect. The full computation of the diffusion coefficients of Fe and O is however not in the scope of the present work, and will be discussed in a separate study.

The transport properties of Fe and O in the different materials found along the whole range of oxygen concentration from pure Fe to the stable iron oxides is of upmost importance to understand the mechanisms of oxidation and reduction of iron and its oxides \cite{Ma2022}. Both are envisioned as key intertwined processes in the production of carbon-free energy from combustion of iron/iron oxides particles and the purification of iron through hydrogen-based direct reduction of iron oxides.\\


\begin{figure*}[!htb]    
    \hspace{0mm}
    \includegraphics[trim = 0mm 0mm 0mm 0mm, clip, width=1\linewidth]{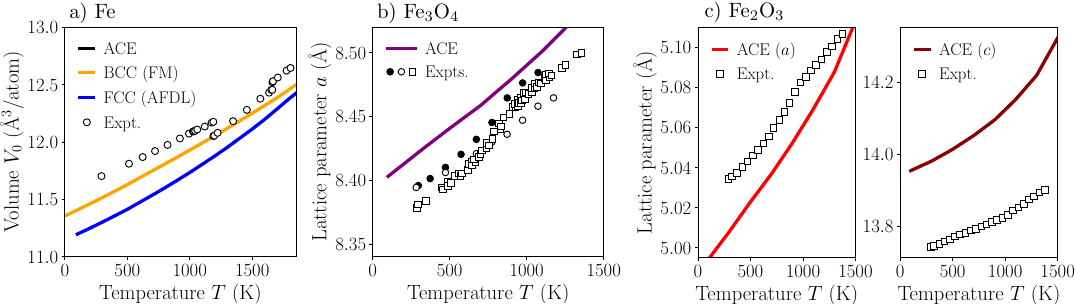}
    \caption{Thermal expansion computed using the ACE potential during molecular dynamics simulations at constant zero pressure for a) pure Fe (BCC and FCC), b) {\magnetite} and c) {\hematite}. Results are compared to experimental data taken from Ref. \cite{Basinski1955} for Fe, from Ref. \cite{Levy2012} (circles) and Ref. \cite{Arkharov1972} (squares) for {\magnetite}, and from Refs. \cite{Saito1965,Neskovic1977} for {\hematite}. For {\hematite}, the two $a$ (left) and $c$ (right) lattice parameters of the hexagonal cell are presented.}
    \label{fig:thermal_exp}
\end{figure*}


\noindent
\textbf{Applications}

\noindent
\textit{Thermal expansion}

\noindent
In view of the targeted applications, it is also important to ensure that the ACE potential is able to capture finite temperature properties of the materials of interest. As in previous sections, we focus on pure Fe and the two iron oxides {\magnetite} and {\hematite}. We present on Fig. \ref{fig:thermal_exp} the lattice expansion of these three prototypes, obtained by NPT molecular dynamics simulations under zero pressure. The magnetic order is kept fixed to the 0\,K ground-state of each material, namely FM and AFDL (anti-ferromagnetic double layer) for BCC and FCC Fe respectively, ferrimagnetic for {\magnetite}, and AF for {\hematite} (see Methods for details). 

Comparing to experimental data, we note that the potential is able to accurately describe the lattice expansion of pure Fe (Fig. \ref{fig:thermal_exp}\,a), the underestimation predicted by ACE being caused by the overbinding of the GGA-PBE functional. As for {\magnetite} and {\hematite} (Figs. \ref{fig:thermal_exp}.b and c respectively), we note that ACE captures well the trend in the increase of lattice parameter of both oxides compared to experimental data. For {\hematite}, the potential slightly overestimates the $c$ lattice parameter of the hexagonal cell, but still reproduces the sharp increase observed experimentally. The slight changes in slope observed in the experimental references for the two oxides correspond to the Néel transition temperature from ordered magnetic state to disordered PM state (around 950\,K for {\magnetite} and {\hematite} \cite{Cornell2003}). This is not accounted for in the results presented on Fig. \ref{fig:thermal_exp}, since the magnetic orders have been kept fixed to the 0\,K ground-state for all materials.


\begin{figure*}[!htb]    
    \includegraphics[trim = 0mm 0mm 0mm 0mm, clip, width=1\linewidth]{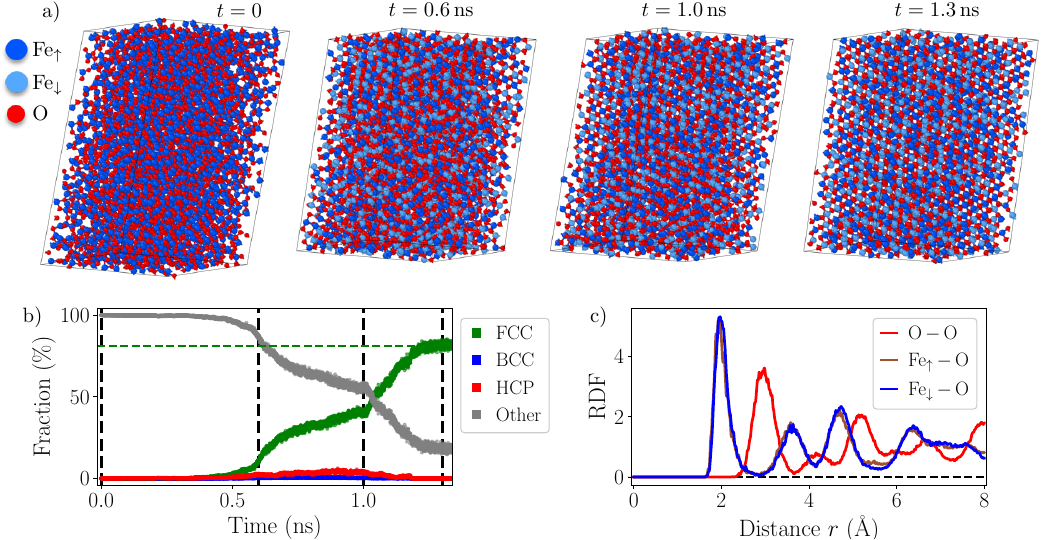}
    \caption{a) Snapshots of an annealing MD-MC simulation of an initially random structure comprising 4480 atoms and a fixed oxygen concentration $x_{\text{O}}=55\%$. b) Evolution of the symmetry of the underlying oxygen sub-lattice as a function of simulation time obtained using a common neighbor analysis algorithm as implemented in the \textsc{Ovito} software \cite{Stukowski2009}. The times of the snapshots shown in a) are marked by vertical lines. c) Radial distribution function (RDF) for pairs of Fe$_{\uparrow}\,-\,$O (brown), Fe$_{\downarrow}\,-\,$O (purple) and O$\,-\,$O (red) after annealing the system for 1.3\,ns.}
    \label{fig:anneal}
\end{figure*}

ACE predicts BCC Fe to melt at approximately 2250\,K when the magnetic order is kept fixed to the FM state, which is overestimated with respect to the experimental value of about 1800\,K \cite{Stull1971}. As for {\magnetite} and {\hematite}, they are predicted to melt at approximately 1750\,K and 1650\,K respectively, which compares quite well with the experimental values of 1870\,K and 1735\,K \cite{Stull1971}. It is also worth noting that the melting point of these two materials constitutes the major identified caveat of the ABOP2019 potential, predicting {\magnetite} and {\hematite} to melt already at room temperature.\\



\noindent
\textit{Annealing of initially random structures}

\noindent
We present in this section a type of tests designed to assess both the robustness of the presented ACE potential and its ability to describe the thermodynamic properties of the Fe-O system through annealing simulations.

Starting from an initially random periodic structure containing a given concentration of oxygen $x_{\text{O}}$, the system is annealed at constant temperature and pressure during a NPT molecular dynamics run. During the annealing, Fe atoms are allowed to change their spin state through semi-grand canonical Monte Carlo swaps performed between the different fixed spin species the ACE potential can handle (see Methods for details). In the example presented on Fig. \ref{fig:anneal}, the structure consists of 4480 atoms with $x_{\text{O}}=55\%$ annealed at 800\,K under zero applied pressure.

After a few steps, the system finds a way out of the highly non-equilibrium starting configuration, and soon starts to order, demonstrating the robustness of the potential, but also its ability to describe high-energy structures which can be found, e.g., in the liquid phase. During these simulations, we kept track of the extrapolation grade $\gamma$ \cite{Lysogorskiy2023} to evaluate uncertainty and to estimate extrapolation of the potential. Apart from the very first steps of the simulation, the potential interpolates with a maximum $\gamma \leq 1$.

According to the experimental phase diagram (see Fig. \ref{fig:ref}\,a), at an oxygen concentration of $55\,\%$ and a temperature of 800\,K, the equilibrium structure should consist of a mixture of wüstite ({\wustite}) and {\magnetite}. To determine the nature of the system after annealing, a simple descriptor to discriminate between different oxide structures is the symmetry of the underlying oxygen sub-lattice. Indeed, for wüstite and {\magnetite}, the oxygen atoms arrange on a FCC lattice, while for {\hematite}, the oxygen sub-lattice is HCP \cite{Cornell2003}. Under the above conditions, we thus expect a FCC O sub-lattice to form. As presented in Fig. \ref{fig:anneal}\,b, the fraction of O atoms with a FCC environment gradually increases as the system is annealed, showing the system to anneal towards a mixture of {\wustite} and {\magnetite}.

The radial distribution functions for Fe-O pairs, seen on Fig. \ref{fig:anneal}\,c, show a peak at around 2\,{\AA} corresponding to the nearest neighbor distance between Fe and O atoms. This is close to its averaged value between 2.1\,{\AA} in wüstite and 1.9\,{\AA}/2.1\,{\AA} in {\magnetite} (for tetrahedrally and octahedrally coordinated Fe atoms respectively) under the same conditions of temperature and pressure. These observations demonstrate the ability of the ACE potential to describe a wide range of environments in terms of both oxygen concentration and atomic structures, while accurately capturing the thermodynamic properties of the Fe-O system.\\



\noindent
\textit{Surface and interface decohesion}


\begin{figure}[!htb]
    \hspace{0mm}
    \includegraphics[trim = 0mm 0mm 0mm 0mm, clip, width=0.925\linewidth]{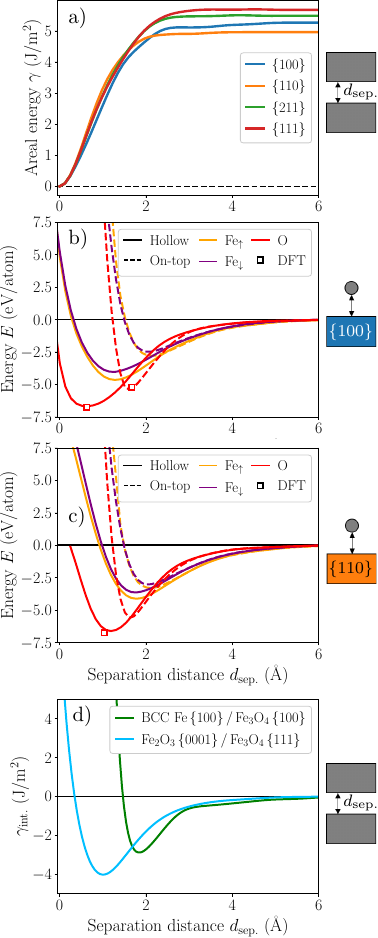}
    \caption{Decohesion in various environments: a)-c) BCC Fe surfaces and adatoms, and d) interfaces. a) Rigid decohesion of various surface facets of BCC Fe. Adatom detachment on a b) \hkl{100} and c) \hkl{110} surfaces of BCC Fe. DFT references for O adsorption are taken from Ref. \cite{Eder2001}. d) Rigid decohesion of BCC Fe/{\magnetite} and {\hematite}/{\magnetite} coherent interfaces.}
    \label{fig:decohesion}
\end{figure}

\noindent
We present in Fig. \ref{fig:decohesion} several decohesion scenarios involving different types of bonds and bonding environments (surfaces, adatoms and mixed interfaces). These pose stringent tests for any potential and several MLIPs fail to predict reasonable decohesion curves, compared to more simple and physically justified interatomic potentials \cite{Lysogorskiy2021}. These tests also have practical applications, for instance in the context of fracture, surface adsorption and interface cohesion in the cases examined here. We stress that none of these tests were explicitly included in the training database, except the equilibrium configuration of the Fe/{\magnetite} interface discussed below.

The rigid decohesion of the four low-index surfaces of BCC Fe (\hkl{100}, \hkl{110}, \hkl{211} and \hkl{111}) presented in Fig. \ref{fig:decohesion}\,a show a smooth behavior, converging to twice the surface energy of each facet (see Supplementary Materials for comparison). The \hkl{110} surface has the lowest energy, closely followed by \hkl{100}, as also reported in previous DFT studies \cite{Dragoni2018,Rinaldi2024}. Now considering adatom adsorption on BCC Fe surfaces (see Fig. \ref{fig:decohesion}\,b and c for \hkl{100} and \hkl{110} facets respectively), the adhesion profiles at the high-symmetry hollow and on-top sites as a function of the separation distance to the surface show a smooth behavior for all atomic species supported by the present ACE potential (Fe, both spin up and down, and O). Most particularly, the adsorption energy of an O atom on the two surfaces of BCC Fe predicted by ACE agree very well with previous DFT studies on both \hkl{100} and \hkl{110} surfaces \cite{Eder2001,Blonski2005}. This property has been considered as one of the benchmarks for assessing the accuracy of various reaxFF potentials in a recent work by Thijs {\etal} \cite{Thijs2023}, where the tested parameterizations yielded a considerable spread in the predicted adsorption energies compared to DFT.

We finally focus on bond-breaking environments at mixed coherent interfaces, {\ie} between pure Fe and its oxides, or between different iron oxides. Such interfaces are fundamental in the understanding of dynamic processes such as hydrogen reduction, and more generally bulk phase transformations, involving coexistence of multiple phases \cite{Zhang2023}. Modeling of these complex structures requires an accurate description of bulk environments in both materials, but also highly distorted structures found at the very interface. We considered two different interfaces: BCC Fe/{\magnetite} and {\magnetite}/{\hematite}.

The orientation relationships of the two interfaces were both determined experimentally in previous works. For the BCC Fe/{\magnetite} interface, it corresponds to Fe\hkl[001]$_{z}\parallel\,${\magnetite}\hkl[001]$_z$ and Fe\hkl[100]$_{x}\parallel\,${\magnetite}\hkl[110]$_x$ \cite{Davenport2000}. For the {\magnetite}/{\hematite} interface, the orientation relationship corresponds to {\magnetite}\hkl[111]$_{z}\parallel\,${\hematite}\hkl[0001]$_z$ and {\magnetite}\hkl[11-2]$_{x}\parallel\,${\hematite}\hkl[11-20]$_x$ \cite{Zhang2023}. In the BCC Fe/{\magnetite} interface, the relative position of the two slabs yielding the most stable configuration of the interface corresponds to the Fe atoms of the outer {\magnetite}\hkl{100} surfaces matching the hollow sites of the BCC Fe\hkl{100} surfaces \cite{Zhou2024_preprint}. For the {\magnetite}/{\hematite} interface, the relative position of the two slabs corresponds to the position in which the two O sub-lattices of {\magnetite} (FCC) and {\hematite} (HCP) match such, that the outermost layers of {\magnetite} coincides with the HCP lattice of {\hematite} \cite{Zhang2023}.

The decohesion curves giving the work of adhesion $\gamma_{\text{int.}}$ of the interfaces are presented in Fig. \ref{fig:decohesion}\,d, showing a smooth variation as a function of the separation distance between the two slabs. In a previous work focusing on the Fe/{\magnetite} interface \cite{Zhou2024_preprint}, we reported an adhesion energy and equilibrium separation distance of $-2.96$\,J/m$^2$ and $1.9$\,{\AA}, respectively, within DFT, which agree very well with the $-2.88$\,J/m$^2$ and $1.8$\,{\AA} values predicted by the ACE potential. The adhesion of the {\magnetite}/{\hematite} interface is predicted to be stronger than for the Fe/{\magnetite} interface, with a shorter separation distance of $1.00$\,{\AA}. This is due to the strong binding energy obtained from merging the O sub-lattices of the two oxides. Additionally, since most of the previous interatomic potentials fitted for the Fe-O cannot accurately describe both pure Fe and every oxide structure, we demonstrate the present ACE potential to be currently the only one able to study these extended defects.

More details about the setup and the geometry of the interface configurations are given in Methods and in the Supplementary Information to this work.\\




\noindent
\textsc{\textbf{Discussion}}

\noindent
In the present work, we proposed an accurate and transferable ACE interatomic potential able to describe the complexity and variety encountered within the Fe-O system across the whole range of oxygen concentration. Based on an extensive DFT-computed training set encompassing structures spanning a wide range of structural and chemical environments, we demonstrate the robustness of the fitted ACE potential on a variety of properties of interest for the many applications iron and its oxides are considered for. The focus was put mainly on thermodynamics, the point defect properties, the adsorption and finite temperature bulk properties, for which we report a very good agreement with DFT reference and experimental data. Additionally, we demonstrated the ability of the potential to extrapolate to random structures at larger length and time scales.

The model overcomes the limitations of pre-existing potentials, mainly based on the reaxFF formalism, when it comes to describing solid phases of the Fe-O binary system \cite{Thijs2023}. Given the variety of the DFT data underlying its parameterization, we also expect the potential to perform reasonably well for the liquid phases of the Fe-O system, which is important in the context of combustion.

Explicit inclusion of magnetism in the proposed Ising-like manner also allows to perform large scale simulations where both magnetic and atomic degrees of freedom can evolve dynamically at a reasonable computational cost. Moreover, discretization of magnetic moments allows us to avoid the expensive sampling of magnetic moment magnitudes, which is necessary for more flexible models \cite{Rinaldi2024}, thus enhancing data efficiency.

However, there exist some caveats to the present ACE potential. Keeping in mind that a MLIP can only be as "good" and trustworthy as the DFT used to compute the data for its training, in the case of the Fe-O system this is a challenge, since it is not clear which DFT "flavor" is suitable to describe both metallic Fe and its insulating oxides equally well on the same footing. Consequences of our choice to use the standard GGA-PBE functional can be seen throughout the results presented in this work. For instance, the experimentally reported high pressure and temperature phase \cite{Hu2016}  FeO$_2$ (see Fig. \ref{fig:ref}\,b) is predicted as not being stable by {\dftu}. Also, stoichiometric wüstite (FeO) is not thermodynamically stable at any temperature under zero pressure, but dissociates into a mixture of Fe and {\magnetite}. At finite pressures, FeO is however stable using the present ACE potential. 


Still, since the GGA-PBE functional yields a good agreement with experiments in terms of structural properties, and since the ACE potential agrees very well with the DFT reference data, we expect the potential not to yield unreliable results when it comes to predicting the stability of various structures. 

Regarding the description of magnetism, an Ising-like model as proposed in the present work obviously does not allow for accurate modeling of magnetic excitations. A proper account of these excitations is required to predict, for instance, the paramagnetic transition from BCC to FCC Fe, which a carefully parameterized ACE potential including non-collinear effects ({\ie} both variations in the magnitude and orientation of magnetic moments) was recently demonstrated to capture \cite{Rinaldi2024}. Nonetheless, even if using this rather crude description, the potential is able to capture most of the important features of magnetism for both pure Fe and its oxides. Similarly, the explicit account of charges might also be of great importance to capture long-range interactions in the oxides.

Overall, we believe the presented ACE model to be the first interatomic potential able to reliably describe the entire range of oxygen concentration from pure iron to its oxides, enabling the access of length and time scales otherwise inaccessible by DFT calculations.\\

\noindent
\textsc{\textbf{METHODS}}

\noindent
\textbf{DFT calculations}

\noindent
All DFT calculations presented in the work were carried out using the \textsc{VASP} package \cite{Kresse1996}. As discussed in the main text, the GGA-PBE xc-functional \cite{Perdew1996} was used in all calculations. Projector-augmented wave (PAW) pseudo-potentials \cite{Blochl1994} were used to describe Fe and O atoms, including 8 and 6 valence electrons respectively. We use a plane-wave energy cutoff of 500\,eV for all DFT calculations, with a $\Gamma$-centered $k$-point mesh of density $0.02\,2\,\pi$\,{\AA}$^{-1}$. Magnetism is included in all calculations in its collinear approximation within spin-polarized DFT.

When performed, atomic relaxations are considered converged when the maximum component of the remaining forces on all atoms is less than 5\,meV/{\AA}.\\

\noindent
\textbf{Diffusion barriers}

\noindent
Vacancy and interstitial migration barriers (see Tab. \ref{tab:diff_barriers_bccFe} and Fig. \ref{fig:diff_ox}) were computed using the nudged elastic band (NEB) method considering five intermediate images linked by a spring constant of 5\,meV/{\AA}. Single vacancy and oxygen migration barriers in BCC Fe presented in Tab. \ref{tab:diff_barriers_bccFe} were computed in a $3\times3\times3$ supercell containing 54 atoms. The vacancy migration barriers in {\magnetite} presented in Fig. \ref{fig:diff_ox}\,a were computed in the conventional cubic cell containing 56 atoms. Vacancy migration barriers in {\hematite} presented in Fig. \ref{fig:diff_ox}\,b were computed in a rhombohedral cell having axes $x \parallel \hkl[1-100]$, $y \parallel \hkl[11-20]$ and $z \parallel \hkl[0001]$ containing 120 atoms.\\

\noindent
\textbf{MD-MC simulations}

\noindent
Hybrid molecular dynamics-Monte Carlo (MD-MC) simulations presented in this work have been performed using the \textsc{LAMMPS} code \cite{Thompson2022}. In particular, we use MD-MC simulations to equilibrate the magnetic structure of a structure. To do so, we allow Fe atoms to swap between the two spin up and spin down species handled by the ACE potential. This is performed within a semi-grand canonical MC swap algorithm as implemented in \textsc{LAMMPS}, where the proportion of spin up and down Fe atoms is allowed to change along the simulation \cite{Sadigh2012}. The chemical potential difference between the two spin states while performing the swapping attempts is set to zero, and the temperature for the MC swaps is set to the temperature of the ongoing MD. A timestep of 1\,fs was used for all MD simulations performed in this work. We perform 100 MC swap attempts every 10 steps of the MD. The initial structure used in the annealing simulation presented in Fig. \ref{fig:anneal} was generated using the \textsc{Buildcell} code from the AIRSS suite \cite{Pickard2011}.\\

\noindent
\textbf{Thermal expansion}

\noindent
The lattice expansion of Fe (BCC and FCC phases), {\magnetite} and {\hematite} presented in Fig. \ref{fig:thermal_exp} were obtained through averaging the lattice parameters of the materials over NPT molecular dynamics runs under zero pressure during 100\,ps after an initial thermalization step of 10\,ps. The supercells used for BCC and FCC Fe contain 2000 and 2304 atoms respectively. For the two {\magnetite} and {\hematite} iron oxides, the supercells used contain 1512 and 1800 atoms respectively. During the simulation, the magnetic order of each material is kept fixed to its 0\,K ground-state as defined in the main text.

Melting points are estimated as the point in temperature where the variation of the volume as a function of the temperature changes slope abruptly following the NPT procedure described above.\\

\noindent
\textbf{Defect formation energies}

\noindent
For point defects formation energies $E_f$ as a function of the oxygen chemical potential change $\Delta \mu_{\text{O}}$, the following equation was used
\begin{equation}
    E_f=E_{\text{def.}}-E_{\text{bulk}} \pm (E_{\text{ref.}}^{i} + \Delta \mu_i),
\end{equation}
where $E_{\text{def.}}$ and $E_{\text{bulk}}$ are the total energies of the simulation cell containing a point defect and the perfect bulk crystal respectively, and $\mu_i = E_{\text{ref.}}^{i} + \Delta \mu_i$ is the chemical potential of species $i$. A plus (minus) sign applies for vacancies (interstitials). 

We impose Fe and O atoms to be in equilibrium with the material considered, enforcing the relation $E_{\text{Fe}_3\text{O}_4}=3\,\mu_{\text{Fe}} + 4\,\mu_{\text{O}}$ in {\magnetite} and resulting in $\mu_{\text{Fe}}=\sfrac{1}{3}\left[ E_{\text{Fe}_3\text{O}_4} - 4\,\mu_{\text{O}} \right]$. Similarly in {\hematite}, we have $E_{\text{Fe}_2\text{O}_3}=2\,\mu_{\text{Fe}} + 3\,\mu_{\text{O}}$, and thus $\mu_{\text{Fe}}=\sfrac{1}{2}\left[ E_{\text{Fe}_2\text{O}_3} - 3\,\mu_{\text{O}} \right]$. These relations then allow to plot the formation energies of point defects as a function of the oxygen chemical potential only, as presented in Fig. \ref{fig:muO_defects}. The convergence of the formation energies with respect to the size of the simulation cell was checked for each prototype material presented.\\

\noindent
\textbf{Interface structures}

\noindent
For the two interfaces presented in Fig. \ref{fig:decohesion}\,d, we use slab structures periodic in the two in-plane directions with open surfaces and a 20\,{\AA}-thick vacuum layer in the direction normal to the interface plane. The thickness of the two slabs in the direction normal to the interface is set to approximately 20\,{\AA}. The outermost atomic layers in the direction normal to the interface are kept fixed during all subsequent calculations.

Before evaluating the interfacial energies of the two cases, we first searched for the relative position of the two slabs that minimizes the total energy of the interface structures, which are described in the main text. The distance between the two slabs in the direction normal to the interface plane is then varied, and the interfacial energy $\gamma_{\text{int.}}$ is evaluated as
\begin{equation}
    \gamma_{\text{int.}}=\Big[ E^{\text{tot.}} - \big( E^{\text{slab 1}} + E^{\text{slab 2}} \big) \Big]/S,
\end{equation}
with $E^{\text{tot.}}$ the total energy of the simulation cell containing the two slabs of energies $E^{\text{slab 1}}$ and $E^{\text{slab 2}}$, and $S$ the surface of the interface structure.\\

\noindent
\textsc{\textbf{Data availability}}

\noindent
The authors declare that all data supporting the findings of the present study are available from B.B. upon reasonale request.\\

\noindent
\textsc{\textbf{Competing interests}}

\noindent
The authors declare no competing interests.\\

\noindent
\textsc{\textbf{Acknowledgments}}

\noindent
We acknowledge funding by the Deutsche Forschungsgemeinschaft (DFG, German Research Foundation) through Project No. 409476157 (SFB1394). B.B also acknowledges support from the Alexander von Humboldt Foundation-Stiftung. D.R. acknowledges funding by the European Union, through the project ROC, sponsored by the European Research Council (ERC, Grant No. 101054368).


\bibliographystyle{apsrev4-2}
\bibliography{biblio}

\begin{thebibliography}{65}%
\makeatletter
\providecommand \@ifxundefined [1]{%
 \@ifx{#1\undefined}
}%
\providecommand \@ifnum [1]{%
 \ifnum #1\expandafter \@firstoftwo
 \else \expandafter \@secondoftwo
 \fi
}%
\providecommand \@ifx [1]{%
 \ifx #1\expandafter \@firstoftwo
 \else \expandafter \@secondoftwo
 \fi
}%
\providecommand \natexlab [1]{#1}%
\providecommand \enquote  [1]{``#1''}%
\providecommand \bibnamefont  [1]{#1}%
\providecommand \bibfnamefont [1]{#1}%
\providecommand \citenamefont [1]{#1}%
\providecommand \href@noop [0]{\@secondoftwo}%
\providecommand \href [0]{\begingroup \@sanitize@url \@href}%
\providecommand \@href[1]{\@@startlink{#1}\@@href}%
\providecommand \@@href[1]{\endgroup#1\@@endlink}%
\providecommand \@sanitize@url [0]{\catcode `\\12\catcode `\$12\catcode
  `\&12\catcode `\#12\catcode `\^12\catcode `\_12\catcode `\%12\relax}%
\providecommand \@@startlink[1]{}%
\providecommand \@@endlink[0]{}%
\providecommand \url  [0]{\begingroup\@sanitize@url \@url }%
\providecommand \@url [1]{\endgroup\@href {#1}{\urlprefix }}%
\providecommand \urlprefix  [0]{URL }%
\providecommand \Eprint [0]{\href }%
\providecommand \doibase [0]{https://doi.org/}%
\providecommand \selectlanguage [0]{\@gobble}%
\providecommand \bibinfo  [0]{\@secondoftwo}%
\providecommand \bibfield  [0]{\@secondoftwo}%
\providecommand \translation [1]{[#1]}%
\providecommand \BibitemOpen [0]{}%
\providecommand \bibitemStop [0]{}%
\providecommand \bibitemNoStop [0]{.\EOS\space}%
\providecommand \EOS [0]{\spacefactor3000\relax}%
\providecommand \BibitemShut  [1]{\csname bibitem#1\endcsname}%
\let\auto@bib@innerbib\@empty
\bibitem [{\citenamefont {Cornell}\ and\ \citenamefont
  {Schwertmann}(2003)}]{Cornell2003}%
  \BibitemOpen
  \bibfield  {author} {\bibinfo {author} {\bibfnamefont {R.~N.}\ \bibnamefont
  {Cornell}}\ and\ \bibinfo {author} {\bibfnamefont {U.}~\bibnamefont
  {Schwertmann}},\ }\href {https://doi.org/10.1002/3527602097} {\emph {\bibinfo
  {title} {The Iron Oxides: Structure, Properties, Reactions, Occurences and
  Uses}}}\ (\bibinfo  {publisher} {Wiley‐VCH Verlag GmbH \& Co. KGaA},\
  \bibinfo {year} {2003})\BibitemShut {NoStop}%
\bibitem [{\citenamefont {Meng}\ \emph {et~al.}(2016)\citenamefont {Meng},
  \citenamefont {Liu}, \citenamefont {Huo}, \citenamefont {Guo}, \citenamefont
  {Cao}, \citenamefont {Peng}, \citenamefont {Dearden}, \citenamefont {Gonze},
  \citenamefont {Yang}, \citenamefont {Wang}, \citenamefont {Jiao},
  \citenamefont {Li},\ and\ \citenamefont {Wen}}]{Meng2016}%
  \BibitemOpen
  \bibfield  {author} {\bibinfo {author} {\bibfnamefont {Y.}~\bibnamefont
  {Meng}}, \bibinfo {author} {\bibfnamefont {X.-W.}\ \bibnamefont {Liu}},
  \bibinfo {author} {\bibfnamefont {C.-F.}\ \bibnamefont {Huo}}, \bibinfo
  {author} {\bibfnamefont {W.-P.}\ \bibnamefont {Guo}}, \bibinfo {author}
  {\bibfnamefont {D.-B.}\ \bibnamefont {Cao}}, \bibinfo {author} {\bibfnamefont
  {Q.}~\bibnamefont {Peng}}, \bibinfo {author} {\bibfnamefont {A.}~\bibnamefont
  {Dearden}}, \bibinfo {author} {\bibfnamefont {X.}~\bibnamefont {Gonze}},
  \bibinfo {author} {\bibfnamefont {Y.}~\bibnamefont {Yang}}, \bibinfo {author}
  {\bibfnamefont {J.}~\bibnamefont {Wang}}, \bibinfo {author} {\bibfnamefont
  {H.}~\bibnamefont {Jiao}}, \bibinfo {author} {\bibfnamefont {Y.}~\bibnamefont
  {Li}},\ and\ \bibinfo {author} {\bibfnamefont {X.-D.}\ \bibnamefont {Wen}},\
  }\href {https://doi.org/10.1021/acs.jctc.6b00640} {\bibfield  {journal}
  {\bibinfo  {journal} {J. Chem. Theory Comput.}\ }\textbf {\bibinfo {volume}
  {12}},\ \bibinfo {pages} {5132} (\bibinfo {year} {2016})}\BibitemShut
  {NoStop}%
\bibitem [{\citenamefont {Chenoweth}\ \emph {et~al.}(2008)\citenamefont
  {Chenoweth}, \citenamefont {van Duin},\ and\ \citenamefont
  {Goddard}}]{Chenoweth2008}%
  \BibitemOpen
  \bibfield  {author} {\bibinfo {author} {\bibfnamefont {K.}~\bibnamefont
  {Chenoweth}}, \bibinfo {author} {\bibfnamefont {A.~C.~T.}\ \bibnamefont {van
  Duin}},\ and\ \bibinfo {author} {\bibfnamefont {W.~A.}\ \bibnamefont
  {Goddard}},\ }\href {https://doi.org/10.1021/jp709896w} {\bibfield  {journal}
  {\bibinfo  {journal} {J. Phys. Chem. A}\ }\textbf {\bibinfo {volume} {112}},\
  \bibinfo {pages} {1040} (\bibinfo {year} {2008})}\BibitemShut {NoStop}%
\bibitem [{\citenamefont {Thijs}\ \emph {et~al.}(2023)\citenamefont {Thijs},
  \citenamefont {Kritikos}, \citenamefont {Giusti}, \citenamefont {van Ende},
  \citenamefont {van Duin},\ and\ \citenamefont {Mi}}]{Thijs2023}%
  \BibitemOpen
  \bibfield  {author} {\bibinfo {author} {\bibfnamefont {L.~C.}\ \bibnamefont
  {Thijs}}, \bibinfo {author} {\bibfnamefont {E.~M.}\ \bibnamefont {Kritikos}},
  \bibinfo {author} {\bibfnamefont {A.}~\bibnamefont {Giusti}}, \bibinfo
  {author} {\bibfnamefont {M.-A.}\ \bibnamefont {van Ende}}, \bibinfo {author}
  {\bibfnamefont {A.~C.~T.}\ \bibnamefont {van Duin}},\ and\ \bibinfo {author}
  {\bibfnamefont {X.}~\bibnamefont {Mi}},\ }\href
  {https://doi.org/10.1021/acs.jpca.3c06646} {\bibfield  {journal} {\bibinfo
  {journal} {J. Phys. Chem. A}\ }\textbf {\bibinfo {volume} {127}},\ \bibinfo
  {pages} {10339} (\bibinfo {year} {2023})}\BibitemShut {NoStop}%
\bibitem [{\citenamefont {Aryanpour}\ \emph {et~al.}(2010)\citenamefont
  {Aryanpour}, \citenamefont {van Duin},\ and\ \citenamefont
  {Kubicki}}]{Aryanpour2010}%
  \BibitemOpen
  \bibfield  {author} {\bibinfo {author} {\bibfnamefont {M.}~\bibnamefont
  {Aryanpour}}, \bibinfo {author} {\bibfnamefont {A.~C.~T.}\ \bibnamefont {van
  Duin}},\ and\ \bibinfo {author} {\bibfnamefont {J.~D.}\ \bibnamefont
  {Kubicki}},\ }\href {https://doi.org/10.1021/jp101332k} {\bibfield  {journal}
  {\bibinfo  {journal} {J. Phys. Chem. A}\ }\textbf {\bibinfo {volume} {114}},\
  \bibinfo {pages} {6298} (\bibinfo {year} {2010})}\BibitemShut {NoStop}%
\bibitem [{\citenamefont {Byggmästar}\ \emph {et~al.}(2019)\citenamefont
  {Byggmästar}, \citenamefont {Nagel}, \citenamefont {Albe}, \citenamefont
  {Henriksson},\ and\ \citenamefont {Nordlund}}]{Byggmastar2019}%
  \BibitemOpen
  \bibfield  {author} {\bibinfo {author} {\bibfnamefont {J.}~\bibnamefont
  {Byggmästar}}, \bibinfo {author} {\bibfnamefont {M.}~\bibnamefont {Nagel}},
  \bibinfo {author} {\bibfnamefont {K.}~\bibnamefont {Albe}}, \bibinfo {author}
  {\bibfnamefont {K.~O.~E.}\ \bibnamefont {Henriksson}},\ and\ \bibinfo
  {author} {\bibfnamefont {K.}~\bibnamefont {Nordlund}},\ }\href
  {https://doi.org/10.1088/1361-648X/ab0931} {\bibfield  {journal} {\bibinfo
  {journal} {J. Phys. Cond. Mat.}\ }\textbf {\bibinfo {volume} {31}},\ \bibinfo
  {pages} {215401} (\bibinfo {year} {2019})}\BibitemShut {NoStop}%
\bibitem [{\citenamefont {Drautz}(2019)}]{Drautz2019}%
  \BibitemOpen
  \bibfield  {author} {\bibinfo {author} {\bibfnamefont {R.}~\bibnamefont
  {Drautz}},\ }\bibfield  {journal} {\bibinfo  {journal} {Phys. Rev. B}\
  }\textbf {\bibinfo {volume} {99}},\ \href
  {https://doi.org/10.1103/PhysRevB.99.014104} {10.1103/PhysRevB.99.014104}
  (\bibinfo {year} {2019})\BibitemShut {NoStop}%
\bibitem [{\citenamefont {Darken}\ and\ \citenamefont
  {Gurry}(1946)}]{Darken1946}%
  \BibitemOpen
  \bibfield  {author} {\bibinfo {author} {\bibfnamefont {L.~S.}\ \bibnamefont
  {Darken}}\ and\ \bibinfo {author} {\bibfnamefont {R.~W.}\ \bibnamefont
  {Gurry}},\ }\href {https://doi.org/10.1021/ja01209a030} {\bibfield  {journal}
  {\bibinfo  {journal} {J. Am. Chem. Soc.}\ }\textbf {\bibinfo {volume} {68}},\
  \bibinfo {pages} {798} (\bibinfo {year} {1946})}\BibitemShut {NoStop}%
\bibitem [{\citenamefont {Dorogokupets}\ \emph {et~al.}(2017)\citenamefont
  {Dorogokupets}, \citenamefont {Dymshits}, \citenamefont {Litasov},\ and\
  \citenamefont {Sokolova}}]{Dorogokupets2017}%
  \BibitemOpen
  \bibfield  {author} {\bibinfo {author} {\bibfnamefont {P.~I.}\ \bibnamefont
  {Dorogokupets}}, \bibinfo {author} {\bibfnamefont {A.~M.}\ \bibnamefont
  {Dymshits}}, \bibinfo {author} {\bibfnamefont {K.~D.}\ \bibnamefont
  {Litasov}},\ and\ \bibinfo {author} {\bibfnamefont {T.~S.}\ \bibnamefont
  {Sokolova}},\ }\href {https://doi.org/10.1038/srep41863} {\bibfield
  {journal} {\bibinfo  {journal} {Sci. Rep.}\ }\textbf {\bibinfo {volume}
  {7}},\ \bibinfo {pages} {41863} (\bibinfo {year} {2017})}\BibitemShut
  {NoStop}%
\bibitem [{\citenamefont {Hazen}\ and\ \citenamefont
  {Jeanloz}(1984)}]{Hazen1984}%
  \BibitemOpen
  \bibfield  {author} {\bibinfo {author} {\bibfnamefont {R.~M.}\ \bibnamefont
  {Hazen}}\ and\ \bibinfo {author} {\bibfnamefont {R.}~\bibnamefont
  {Jeanloz}},\ }\href {https://doi.org/10.1029/RG022i001p00037} {\bibfield
  {journal} {\bibinfo  {journal} {Rev. Geophys.}\ }\textbf {\bibinfo {volume}
  {22}},\ \bibinfo {pages} {37} (\bibinfo {year} {1984})}\BibitemShut {NoStop}%
\bibitem [{\citenamefont {Dudarev}\ \emph {et~al.}(1998)\citenamefont
  {Dudarev}, \citenamefont {Botton}, \citenamefont {Savrasov}, \citenamefont
  {Humphreys},\ and\ \citenamefont {Sutton}}]{Dudarev1998}%
  \BibitemOpen
  \bibfield  {author} {\bibinfo {author} {\bibfnamefont {S.~L.}\ \bibnamefont
  {Dudarev}}, \bibinfo {author} {\bibfnamefont {G.~A.}\ \bibnamefont {Botton}},
  \bibinfo {author} {\bibfnamefont {S.~Y.}\ \bibnamefont {Savrasov}}, \bibinfo
  {author} {\bibfnamefont {C.~J.}\ \bibnamefont {Humphreys}},\ and\ \bibinfo
  {author} {\bibfnamefont {A.~P.}\ \bibnamefont {Sutton}},\ }\href
  {https://doi.org/10.1103/PhysRevB.57.1505} {\bibfield  {journal} {\bibinfo
  {journal} {Phys. Rev. B}\ }\textbf {\bibinfo {volume} {57}},\ \bibinfo
  {pages} {1505} (\bibinfo {year} {1998})}\BibitemShut {NoStop}%
\bibitem [{\citenamefont {Shang}\ \emph {et~al.}(2014)\citenamefont {Shang},
  \citenamefont {Fang}, \citenamefont {Wang}, \citenamefont {Guo},
  \citenamefont {Wang}, \citenamefont {Jablonski}, \citenamefont {Du},\ and\
  \citenamefont {Liu}}]{Shang2014}%
  \BibitemOpen
  \bibfield  {author} {\bibinfo {author} {\bibfnamefont {S.~L.}\ \bibnamefont
  {Shang}}, \bibinfo {author} {\bibfnamefont {H.~Z.}\ \bibnamefont {Fang}},
  \bibinfo {author} {\bibfnamefont {J.}~\bibnamefont {Wang}}, \bibinfo {author}
  {\bibfnamefont {C.~P.}\ \bibnamefont {Guo}}, \bibinfo {author} {\bibfnamefont
  {Y.}~\bibnamefont {Wang}}, \bibinfo {author} {\bibfnamefont {P.~D.}\
  \bibnamefont {Jablonski}}, \bibinfo {author} {\bibfnamefont {Y.}~\bibnamefont
  {Du}},\ and\ \bibinfo {author} {\bibfnamefont {Z.~K.}\ \bibnamefont {Liu}},\
  }\href {https://doi.org/10.1016/j.corsci.2014.02.009} {\bibfield  {journal}
  {\bibinfo  {journal} {Corrosion Science}\ }\textbf {\bibinfo {volume} {83}},\
  \bibinfo {pages} {94} (\bibinfo {year} {2014})}\BibitemShut {NoStop}%
\bibitem [{\citenamefont {Perdew}\ \emph {et~al.}(1996)\citenamefont {Perdew},
  \citenamefont {Burke},\ and\ \citenamefont {Ernzerhof}}]{Perdew1996}%
  \BibitemOpen
  \bibfield  {author} {\bibinfo {author} {\bibfnamefont {J.~P.}\ \bibnamefont
  {Perdew}}, \bibinfo {author} {\bibfnamefont {K.}~\bibnamefont {Burke}},\ and\
  \bibinfo {author} {\bibfnamefont {M.}~\bibnamefont {Ernzerhof}},\ }\href
  {https://doi.org/10.1103/PhysRevLett.77.3865} {\bibfield  {journal} {\bibinfo
   {journal} {Phys. Rev. Lett.}\ }\textbf {\bibinfo {volume} {77}},\ \bibinfo
  {pages} {3865} (\bibinfo {year} {1996})}\BibitemShut {NoStop}%
\bibitem [{\citenamefont {Jain}\ \emph {et~al.}(2011)\citenamefont {Jain},
  \citenamefont {Hautier}, \citenamefont {Ong}, \citenamefont {Moore},
  \citenamefont {Fischer}, \citenamefont {Persson},\ and\ \citenamefont
  {Ceder}}]{Jain2011}%
  \BibitemOpen
  \bibfield  {author} {\bibinfo {author} {\bibfnamefont {A.}~\bibnamefont
  {Jain}}, \bibinfo {author} {\bibfnamefont {G.}~\bibnamefont {Hautier}},
  \bibinfo {author} {\bibfnamefont {S.~P.}\ \bibnamefont {Ong}}, \bibinfo
  {author} {\bibfnamefont {C.~J.}\ \bibnamefont {Moore}}, \bibinfo {author}
  {\bibfnamefont {C.~C.}\ \bibnamefont {Fischer}}, \bibinfo {author}
  {\bibfnamefont {K.~A.}\ \bibnamefont {Persson}},\ and\ \bibinfo {author}
  {\bibfnamefont {G.}~\bibnamefont {Ceder}},\ }\href
  {https://doi.org/10.1103/PhysRevB.84.045115} {\bibfield  {journal} {\bibinfo
  {journal} {Phys. Rev. B}\ }\textbf {\bibinfo {volume} {84}},\ \bibinfo
  {pages} {045115} (\bibinfo {year} {2011})}\BibitemShut {NoStop}%
\bibitem [{\citenamefont {Jain}\ \emph {et~al.}(2013)\citenamefont {Jain},
  \citenamefont {Ong}, \citenamefont {Hautier}, \citenamefont {Chen},
  \citenamefont {Richards}, \citenamefont {Dacek}, \citenamefont {Cholia},
  \citenamefont {Gunter}, \citenamefont {Skinner}, \citenamefont {Ceder},\ and\
  \citenamefont {Persson}}]{Jain2013}%
  \BibitemOpen
  \bibfield  {author} {\bibinfo {author} {\bibfnamefont {A.}~\bibnamefont
  {Jain}}, \bibinfo {author} {\bibfnamefont {S.~P.}\ \bibnamefont {Ong}},
  \bibinfo {author} {\bibfnamefont {G.}~\bibnamefont {Hautier}}, \bibinfo
  {author} {\bibfnamefont {W.}~\bibnamefont {Chen}}, \bibinfo {author}
  {\bibfnamefont {W.~D.}\ \bibnamefont {Richards}}, \bibinfo {author}
  {\bibfnamefont {S.}~\bibnamefont {Dacek}}, \bibinfo {author} {\bibfnamefont
  {S.}~\bibnamefont {Cholia}}, \bibinfo {author} {\bibfnamefont
  {D.}~\bibnamefont {Gunter}}, \bibinfo {author} {\bibfnamefont
  {D.}~\bibnamefont {Skinner}}, \bibinfo {author} {\bibfnamefont
  {G.}~\bibnamefont {Ceder}},\ and\ \bibinfo {author} {\bibfnamefont {K.~A.}\
  \bibnamefont {Persson}},\ }\href {https://doi.org/10.1063/1.4812323}
  {\bibfield  {journal} {\bibinfo  {journal} {APL Mat.}\ }\textbf {\bibinfo
  {volume} {1}},\ \bibinfo {pages} {011002} (\bibinfo {year}
  {2013})}\BibitemShut {NoStop}%
\bibitem [{\citenamefont {Samara}\ and\ \citenamefont
  {Giardini}(1969)}]{Samara1969}%
  \BibitemOpen
  \bibfield  {author} {\bibinfo {author} {\bibfnamefont {G.~A.}\ \bibnamefont
  {Samara}}\ and\ \bibinfo {author} {\bibfnamefont {A.~A.}\ \bibnamefont
  {Giardini}},\ }\href {https://doi.org/10.1103/PhysRev.186.577} {\bibfield
  {journal} {\bibinfo  {journal} {Phys. Rev.}\ }\textbf {\bibinfo {volume}
  {186}},\ \bibinfo {pages} {577} (\bibinfo {year} {1969})}\BibitemShut
  {NoStop}%
\bibitem [{\citenamefont {Nešković}\ \emph {et~al.}(1977)\citenamefont
  {Nešković}, \citenamefont {Babić},\ and\ \citenamefont
  {Konstantinović}}]{Neskovic1977}%
  \BibitemOpen
  \bibfield  {author} {\bibinfo {author} {\bibfnamefont {N.~B.}\ \bibnamefont
  {Nešković}}, \bibinfo {author} {\bibfnamefont {B.}~\bibnamefont {Babić}},\
  and\ \bibinfo {author} {\bibfnamefont {J.}~\bibnamefont {Konstantinović}},\
  }\href {https://doi.org/10.1002/pssa.2210410252} {\bibfield  {journal}
  {\bibinfo  {journal} {Phys. Stat. Sol. (a)}\ }\textbf {\bibinfo {volume}
  {41}},\ \bibinfo {pages} {K133} (\bibinfo {year} {1977})}\BibitemShut
  {NoStop}%
\bibitem [{\citenamefont {McCammon}(1992)}]{McCammon1992}%
  \BibitemOpen
  \bibfield  {author} {\bibinfo {author} {\bibfnamefont {C.~A.}\ \bibnamefont
  {McCammon}},\ }\href {https://doi.org/10.1016/0304-8853(92)91612-W}
  {\bibfield  {journal} {\bibinfo  {journal} {J. Mag. Mag. Mater.}\ }\textbf
  {\bibinfo {volume} {104-107}},\ \bibinfo {pages} {1937} (\bibinfo {year}
  {1992})}\BibitemShut {NoStop}%
\bibitem [{\citenamefont {Pereira}\ \emph {et~al.}(2012)\citenamefont
  {Pereira}, \citenamefont {Oliveira},\ and\ \citenamefont
  {Murad}}]{Pereira2012}%
  \BibitemOpen
  \bibfield  {author} {\bibinfo {author} {\bibfnamefont {M.~C.}\ \bibnamefont
  {Pereira}}, \bibinfo {author} {\bibfnamefont {L.~C.~A.}\ \bibnamefont
  {Oliveira}},\ and\ \bibinfo {author} {\bibfnamefont {E.}~\bibnamefont
  {Murad}},\ }\href {https://doi.org/10.1180/claymin.2012.047.3.01} {\bibfield
  {journal} {\bibinfo  {journal} {Clay Min.}\ }\textbf {\bibinfo {volume}
  {47}},\ \bibinfo {pages} {285} (\bibinfo {year} {2012})}\BibitemShut
  {NoStop}%
\bibitem [{\citenamefont {Pickard}\ and\ \citenamefont
  {Needs}(2011)}]{Pickard2011}%
  \BibitemOpen
  \bibfield  {author} {\bibinfo {author} {\bibfnamefont {C.~J.}\ \bibnamefont
  {Pickard}}\ and\ \bibinfo {author} {\bibfnamefont {R.~J.}\ \bibnamefont
  {Needs}},\ }\href {https://doi.org/10.1088/0953-8984/23/5/053201} {\bibfield
  {journal} {\bibinfo  {journal} {J. Phys. Cond. Mat.}\ }\textbf {\bibinfo
  {volume} {23}},\ \bibinfo {pages} {053201} (\bibinfo {year}
  {2011})}\BibitemShut {NoStop}%
\bibitem [{\citenamefont {Poul}\ \emph {et~al.}(2023)\citenamefont {Poul},
  \citenamefont {Huber}, \citenamefont {Bitzek},\ and\ \citenamefont
  {Neugebauer}}]{Poul2023}%
  \BibitemOpen
  \bibfield  {author} {\bibinfo {author} {\bibfnamefont {M.}~\bibnamefont
  {Poul}}, \bibinfo {author} {\bibfnamefont {L.}~\bibnamefont {Huber}},
  \bibinfo {author} {\bibfnamefont {E.}~\bibnamefont {Bitzek}},\ and\ \bibinfo
  {author} {\bibfnamefont {J.}~\bibnamefont {Neugebauer}},\ }\href
  {https://doi.org/10.1103/PhysRevB.107.104103} {\bibfield  {journal} {\bibinfo
   {journal} {Phys. Rev. B}\ }\textbf {\bibinfo {volume} {107}},\ \bibinfo
  {pages} {104103} (\bibinfo {year} {2023})}\BibitemShut {NoStop}%
\bibitem [{\citenamefont {Stull}\ and\ \citenamefont
  {Prophet}(1971)}]{Stull1971}%
  \BibitemOpen
  \bibfield  {author} {\bibinfo {author} {\bibfnamefont {D.}~\bibnamefont
  {Stull}}\ and\ \bibinfo {author} {\bibfnamefont {H.}~\bibnamefont
  {Prophet}},\ }\bibfield  {journal} {\bibinfo  {journal} {NIST}\ }\href
  {https://doi.org/10.6028/NBS.NSRDS.37} {10.6028/NBS.NSRDS.37} (\bibinfo
  {year} {1971})\BibitemShut {NoStop}%
\bibitem [{\citenamefont {Hu}\ \emph {et~al.}(2016)\citenamefont {Hu},
  \citenamefont {Kim}, \citenamefont {Yang}, \citenamefont {Yang},
  \citenamefont {Meng}, \citenamefont {Zhang},\ and\ \citenamefont
  {Mao}}]{Hu2016}%
  \BibitemOpen
  \bibfield  {author} {\bibinfo {author} {\bibfnamefont {Q.}~\bibnamefont
  {Hu}}, \bibinfo {author} {\bibfnamefont {D.~Y.}\ \bibnamefont {Kim}},
  \bibinfo {author} {\bibfnamefont {W.}~\bibnamefont {Yang}}, \bibinfo {author}
  {\bibfnamefont {L.}~\bibnamefont {Yang}}, \bibinfo {author} {\bibfnamefont
  {Y.}~\bibnamefont {Meng}}, \bibinfo {author} {\bibfnamefont {L.}~\bibnamefont
  {Zhang}},\ and\ \bibinfo {author} {\bibfnamefont {H.-K.}\ \bibnamefont
  {Mao}},\ }\href {https://doi.org/10.1038/nature18018} {\bibfield  {journal}
  {\bibinfo  {journal} {Nature}\ }\textbf {\bibinfo {volume} {534}},\ \bibinfo
  {pages} {241} (\bibinfo {year} {2016})}\BibitemShut {NoStop}%
\bibitem [{\citenamefont {Qamar}\ \emph {et~al.}(2023)\citenamefont {Qamar},
  \citenamefont {Mrovec}, \citenamefont {Lysogorskiy}, \citenamefont
  {Bochkarev},\ and\ \citenamefont {Drautz}}]{Qamar2023}%
  \BibitemOpen
  \bibfield  {author} {\bibinfo {author} {\bibfnamefont {M.}~\bibnamefont
  {Qamar}}, \bibinfo {author} {\bibfnamefont {M.}~\bibnamefont {Mrovec}},
  \bibinfo {author} {\bibfnamefont {Y.}~\bibnamefont {Lysogorskiy}}, \bibinfo
  {author} {\bibfnamefont {A.}~\bibnamefont {Bochkarev}},\ and\ \bibinfo
  {author} {\bibfnamefont {R.}~\bibnamefont {Drautz}},\ }\href
  {https://doi.org/10.1021/acs.jctc.2c01149} {\bibfield  {journal} {\bibinfo
  {journal} {J. Chem. Theory Comput.}\ }\textbf {\bibinfo {volume} {19}},\
  \bibinfo {pages} {5151} (\bibinfo {year} {2023})}\BibitemShut {NoStop}%
\bibitem [{\citenamefont {Rinaldi}\ \emph {et~al.}(2024)\citenamefont
  {Rinaldi}, \citenamefont {Mrovec}, \citenamefont {Bochkarev}, \citenamefont
  {Lysogorskiy},\ and\ \citenamefont {Drautz}}]{Rinaldi2024}%
  \BibitemOpen
  \bibfield  {author} {\bibinfo {author} {\bibfnamefont {M.}~\bibnamefont
  {Rinaldi}}, \bibinfo {author} {\bibfnamefont {M.}~\bibnamefont {Mrovec}},
  \bibinfo {author} {\bibfnamefont {A.}~\bibnamefont {Bochkarev}}, \bibinfo
  {author} {\bibfnamefont {Y.}~\bibnamefont {Lysogorskiy}},\ and\ \bibinfo
  {author} {\bibfnamefont {R.}~\bibnamefont {Drautz}},\ }\href
  {https://doi.org/10.1038/s41524-024-01196-8} {\bibfield  {journal} {\bibinfo
  {journal} {npj Comput. Mater.}\ }\textbf {\bibinfo {volume} {10}},\ \bibinfo
  {pages} {1} (\bibinfo {year} {2024})}\BibitemShut {NoStop}%
\bibitem [{\citenamefont {Bochkarev}\ \emph {et~al.}(2022)\citenamefont
  {Bochkarev}, \citenamefont {Lysogorskiy}, \citenamefont {Menon},
  \citenamefont {Qamar}, \citenamefont {Mrovec},\ and\ \citenamefont
  {Drautz}}]{Bochkarev2022}%
  \BibitemOpen
  \bibfield  {author} {\bibinfo {author} {\bibfnamefont {A.}~\bibnamefont
  {Bochkarev}}, \bibinfo {author} {\bibfnamefont {Y.}~\bibnamefont
  {Lysogorskiy}}, \bibinfo {author} {\bibfnamefont {S.}~\bibnamefont {Menon}},
  \bibinfo {author} {\bibfnamefont {M.}~\bibnamefont {Qamar}}, \bibinfo
  {author} {\bibfnamefont {M.}~\bibnamefont {Mrovec}},\ and\ \bibinfo {author}
  {\bibfnamefont {R.}~\bibnamefont {Drautz}},\ }\href
  {https://doi.org/10.1103/PhysRevMaterials.6.013804} {\bibfield  {journal}
  {\bibinfo  {journal} {Phys. Rev. Mater.}\ }\textbf {\bibinfo {volume} {6}},\
  \bibinfo {pages} {013804} (\bibinfo {year} {2022})}\BibitemShut {NoStop}%
\bibitem [{\citenamefont {Lysogorskiy}\ \emph {et~al.}(2021)\citenamefont
  {Lysogorskiy}, \citenamefont {Oord}, \citenamefont {Bochkarev}, \citenamefont
  {Menon}, \citenamefont {Rinaldi}, \citenamefont {Hammerschmidt},
  \citenamefont {Mrovec}, \citenamefont {Thompson}, \citenamefont {Csányi},
  \citenamefont {Ortner},\ and\ \citenamefont {Drautz}}]{Lysogorskiy2021}%
  \BibitemOpen
  \bibfield  {author} {\bibinfo {author} {\bibfnamefont {Y.}~\bibnamefont
  {Lysogorskiy}}, \bibinfo {author} {\bibfnamefont {C.~v.~d.}\ \bibnamefont
  {Oord}}, \bibinfo {author} {\bibfnamefont {A.}~\bibnamefont {Bochkarev}},
  \bibinfo {author} {\bibfnamefont {S.}~\bibnamefont {Menon}}, \bibinfo
  {author} {\bibfnamefont {M.}~\bibnamefont {Rinaldi}}, \bibinfo {author}
  {\bibfnamefont {T.}~\bibnamefont {Hammerschmidt}}, \bibinfo {author}
  {\bibfnamefont {M.}~\bibnamefont {Mrovec}}, \bibinfo {author} {\bibfnamefont
  {A.}~\bibnamefont {Thompson}}, \bibinfo {author} {\bibfnamefont
  {G.}~\bibnamefont {Csányi}}, \bibinfo {author} {\bibfnamefont
  {C.}~\bibnamefont {Ortner}},\ and\ \bibinfo {author} {\bibfnamefont
  {R.}~\bibnamefont {Drautz}},\ }\href
  {https://doi.org/10.1038/s41524-021-00559-9} {\bibfield  {journal} {\bibinfo
  {journal} {npj Comput. Mater.}\ }\textbf {\bibinfo {volume} {7}},\ \bibinfo
  {pages} {1} (\bibinfo {year} {2021})}\BibitemShut {NoStop}%
\bibitem [{\citenamefont {Wang}\ \emph {et~al.}(2021)\citenamefont {Wang},
  \citenamefont {Kingsbury}, \citenamefont {McDermott}, \citenamefont {Horton},
  \citenamefont {Jain}, \citenamefont {Ong}, \citenamefont {Dwaraknath},\ and\
  \citenamefont {Persson}}]{Wang2021}%
  \BibitemOpen
  \bibfield  {author} {\bibinfo {author} {\bibfnamefont {A.}~\bibnamefont
  {Wang}}, \bibinfo {author} {\bibfnamefont {R.}~\bibnamefont {Kingsbury}},
  \bibinfo {author} {\bibfnamefont {M.}~\bibnamefont {McDermott}}, \bibinfo
  {author} {\bibfnamefont {M.}~\bibnamefont {Horton}}, \bibinfo {author}
  {\bibfnamefont {A.}~\bibnamefont {Jain}}, \bibinfo {author} {\bibfnamefont
  {S.~P.}\ \bibnamefont {Ong}}, \bibinfo {author} {\bibfnamefont
  {S.}~\bibnamefont {Dwaraknath}},\ and\ \bibinfo {author} {\bibfnamefont
  {K.~A.}\ \bibnamefont {Persson}},\ }\href
  {https://doi.org/10.1038/s41598-021-94550-5} {\bibfield  {journal} {\bibinfo
  {journal} {Sci. Rep.}\ }\textbf {\bibinfo {volume} {11}},\ \bibinfo {pages}
  {15496} (\bibinfo {year} {2021})}\BibitemShut {NoStop}%
\bibitem [{\citenamefont {Basinski}\ \emph {et~al.}(1955)\citenamefont
  {Basinski}, \citenamefont {Hume-Rothery},\ and\ \citenamefont
  {Sutton}}]{Basinski1955}%
  \BibitemOpen
  \bibfield  {author} {\bibinfo {author} {\bibfnamefont {Z.~S.}\ \bibnamefont
  {Basinski}}, \bibinfo {author} {\bibfnamefont {W.}~\bibnamefont
  {Hume-Rothery}},\ and\ \bibinfo {author} {\bibfnamefont {A.~L.}\ \bibnamefont
  {Sutton}},\ }\href {https://doi.org/10.1098/rspa.1955.0102} {\bibfield
  {journal} {\bibinfo  {journal} {Proc. R. Soc. London}\ }\textbf {\bibinfo
  {volume} {229}},\ \bibinfo {pages} {459} (\bibinfo {year}
  {1955})}\BibitemShut {NoStop}%
\bibitem [{\citenamefont {Zhang}(2000)}]{Zhang2000}%
  \BibitemOpen
  \bibfield  {author} {\bibinfo {author} {\bibfnamefont {J.}~\bibnamefont
  {Zhang}},\ }\href {https://doi.org/10.1103/PhysRevLett.84.507} {\bibfield
  {journal} {\bibinfo  {journal} {Phys. Rev. Lett.}\ }\textbf {\bibinfo
  {volume} {84}},\ \bibinfo {pages} {507} (\bibinfo {year} {2000})}\BibitemShut
  {NoStop}%
\bibitem [{\citenamefont {Haavik}\ \emph {et~al.}(2000)\citenamefont {Haavik},
  \citenamefont {Stølen}, \citenamefont {Fjellvåg}, \citenamefont
  {Hanfland},\ and\ \citenamefont {Häusermann}}]{Haavik2000}%
  \BibitemOpen
  \bibfield  {author} {\bibinfo {author} {\bibfnamefont {C.}~\bibnamefont
  {Haavik}}, \bibinfo {author} {\bibfnamefont {S.}~\bibnamefont {Stølen}},
  \bibinfo {author} {\bibfnamefont {H.}~\bibnamefont {Fjellvåg}}, \bibinfo
  {author} {\bibfnamefont {M.}~\bibnamefont {Hanfland}},\ and\ \bibinfo
  {author} {\bibfnamefont {D.}~\bibnamefont {Häusermann}},\ }\href
  {https://doi.org/10.2138/am-2000-0413} {\bibfield  {journal} {\bibinfo
  {journal} {American Mineralogist}\ }\textbf {\bibinfo {volume} {85}},\
  \bibinfo {pages} {514} (\bibinfo {year} {2000})}\BibitemShut {NoStop}%
\bibitem [{\citenamefont {Finger}\ and\ \citenamefont
  {Hazen}(1980)}]{Finger1980}%
  \BibitemOpen
  \bibfield  {author} {\bibinfo {author} {\bibfnamefont {L.~W.}\ \bibnamefont
  {Finger}}\ and\ \bibinfo {author} {\bibfnamefont {R.~M.}\ \bibnamefont
  {Hazen}},\ }\href {https://doi.org/10.1063/1.327451} {\bibfield  {journal}
  {\bibinfo  {journal} {J. App. Phys.}\ }\textbf {\bibinfo {volume} {51}},\
  \bibinfo {pages} {5362} (\bibinfo {year} {1980})}\BibitemShut {NoStop}%
\bibitem [{\citenamefont {Banerjee}\ \emph {et~al.}(2023)\citenamefont
  {Banerjee}, \citenamefont {Holby}, \citenamefont {Kohnert}, \citenamefont
  {Srivastava}, \citenamefont {Asta},\ and\ \citenamefont
  {Uberuaga}}]{Banerjee2023}%
  \BibitemOpen
  \bibfield  {author} {\bibinfo {author} {\bibfnamefont {A.}~\bibnamefont
  {Banerjee}}, \bibinfo {author} {\bibfnamefont {E.~F.}\ \bibnamefont {Holby}},
  \bibinfo {author} {\bibfnamefont {A.~A.}\ \bibnamefont {Kohnert}}, \bibinfo
  {author} {\bibfnamefont {S.}~\bibnamefont {Srivastava}}, \bibinfo {author}
  {\bibfnamefont {M.}~\bibnamefont {Asta}},\ and\ \bibinfo {author}
  {\bibfnamefont {B.~P.}\ \bibnamefont {Uberuaga}},\ }\href
  {https://doi.org/10.1088/2516-1075/acd158} {\bibfield  {journal} {\bibinfo
  {journal} {Electronic Structure}\ }\textbf {\bibinfo {volume} {5}},\ \bibinfo
  {pages} {024007} (\bibinfo {year} {2023})}\BibitemShut {NoStop}%
\bibitem [{\citenamefont {De~Schepper}\ \emph {et~al.}(1983)\citenamefont
  {De~Schepper}, \citenamefont {Segers}, \citenamefont {Dorikens-Vanpraet},
  \citenamefont {Dorikens}, \citenamefont {Knuyt}, \citenamefont {Stals},\ and\
  \citenamefont {Moser}}]{Schepper1983}%
  \BibitemOpen
  \bibfield  {author} {\bibinfo {author} {\bibfnamefont {L.}~\bibnamefont
  {De~Schepper}}, \bibinfo {author} {\bibfnamefont {D.}~\bibnamefont {Segers}},
  \bibinfo {author} {\bibfnamefont {L.}~\bibnamefont {Dorikens-Vanpraet}},
  \bibinfo {author} {\bibfnamefont {M.}~\bibnamefont {Dorikens}}, \bibinfo
  {author} {\bibfnamefont {G.}~\bibnamefont {Knuyt}}, \bibinfo {author}
  {\bibfnamefont {L.~M.}\ \bibnamefont {Stals}},\ and\ \bibinfo {author}
  {\bibfnamefont {P.}~\bibnamefont {Moser}},\ }\href
  {https://doi.org/10.1103/PhysRevB.27.5257} {\bibfield  {journal} {\bibinfo
  {journal} {Phys. Rev. B}\ }\textbf {\bibinfo {volume} {27}},\ \bibinfo
  {pages} {5257} (\bibinfo {year} {1983})}\BibitemShut {NoStop}%
\bibitem [{\citenamefont {Barouh}\ \emph {et~al.}(2014)\citenamefont {Barouh},
  \citenamefont {Schuler}, \citenamefont {Fu},\ and\ \citenamefont
  {Nastar}}]{Barouh2014}%
  \BibitemOpen
  \bibfield  {author} {\bibinfo {author} {\bibfnamefont {C.}~\bibnamefont
  {Barouh}}, \bibinfo {author} {\bibfnamefont {T.}~\bibnamefont {Schuler}},
  \bibinfo {author} {\bibfnamefont {C.-C.}\ \bibnamefont {Fu}},\ and\ \bibinfo
  {author} {\bibfnamefont {M.}~\bibnamefont {Nastar}},\ }\href
  {https://doi.org/10.1103/PhysRevB.90.054112} {\bibfield  {journal} {\bibinfo
  {journal} {Phys. Rev. B}\ }\textbf {\bibinfo {volume} {90}},\ \bibinfo
  {pages} {054112} (\bibinfo {year} {2014})}\BibitemShut {NoStop}%
\bibitem [{\citenamefont {Fu}\ \emph {et~al.}(2007)\citenamefont {Fu},
  \citenamefont {Krčmar}, \citenamefont {Painter},\ and\ \citenamefont
  {Chen}}]{Fu2007}%
  \BibitemOpen
  \bibfield  {author} {\bibinfo {author} {\bibfnamefont {C.~L.}\ \bibnamefont
  {Fu}}, \bibinfo {author} {\bibfnamefont {M.}~\bibnamefont {Krčmar}},
  \bibinfo {author} {\bibfnamefont {G.~S.}\ \bibnamefont {Painter}},\ and\
  \bibinfo {author} {\bibfnamefont {X.-Q.}\ \bibnamefont {Chen}},\ }\href
  {https://doi.org/10.1103/PhysRevLett.99.225502} {\bibfield  {journal}
  {\bibinfo  {journal} {Phys. Rev. Lett.}\ }\textbf {\bibinfo {volume} {99}},\
  \bibinfo {pages} {225502} (\bibinfo {year} {2007})}\BibitemShut {NoStop}%
\bibitem [{\citenamefont {Wang}\ \emph {et~al.}(2020)\citenamefont {Wang},
  \citenamefont {Faßbender},\ and\ \citenamefont {Posselt}}]{Wang2020}%
  \BibitemOpen
  \bibfield  {author} {\bibinfo {author} {\bibfnamefont {X.}~\bibnamefont
  {Wang}}, \bibinfo {author} {\bibfnamefont {J.}~\bibnamefont {Faßbender}},\
  and\ \bibinfo {author} {\bibfnamefont {M.}~\bibnamefont {Posselt}},\ }\href
  {https://doi.org/10.1103/PhysRevB.101.174107} {\bibfield  {journal} {\bibinfo
   {journal} {Phys. Rev. B}\ }\textbf {\bibinfo {volume} {101}},\ \bibinfo
  {pages} {174107} (\bibinfo {year} {2020})}\BibitemShut {NoStop}%
\bibitem [{\citenamefont {Fu}\ \emph {et~al.}(2005)\citenamefont {Fu},
  \citenamefont {Torre}, \citenamefont {Willaime}, \citenamefont {Bocquet},\
  and\ \citenamefont {Barbu}}]{Fu2005}%
  \BibitemOpen
  \bibfield  {author} {\bibinfo {author} {\bibfnamefont {C.-C.}\ \bibnamefont
  {Fu}}, \bibinfo {author} {\bibfnamefont {J.~D.}\ \bibnamefont {Torre}},
  \bibinfo {author} {\bibfnamefont {F.}~\bibnamefont {Willaime}}, \bibinfo
  {author} {\bibfnamefont {J.-L.}\ \bibnamefont {Bocquet}},\ and\ \bibinfo
  {author} {\bibfnamefont {A.}~\bibnamefont {Barbu}},\ }\href
  {https://doi.org/10.1038/nmat1286} {\bibfield  {journal} {\bibinfo  {journal}
  {Nature Mat.}\ }\textbf {\bibinfo {volume} {4}},\ \bibinfo {pages} {68}
  (\bibinfo {year} {2005})}\BibitemShut {NoStop}%
\bibitem [{\citenamefont {Barouh}\ \emph {et~al.}(2015)\citenamefont {Barouh},
  \citenamefont {Schuler}, \citenamefont {Fu},\ and\ \citenamefont
  {Jourdan}}]{Barouh2015}%
  \BibitemOpen
  \bibfield  {author} {\bibinfo {author} {\bibfnamefont {C.}~\bibnamefont
  {Barouh}}, \bibinfo {author} {\bibfnamefont {T.}~\bibnamefont {Schuler}},
  \bibinfo {author} {\bibfnamefont {C.-C.}\ \bibnamefont {Fu}},\ and\ \bibinfo
  {author} {\bibfnamefont {T.}~\bibnamefont {Jourdan}},\ }\href
  {https://doi.org/10.1103/PhysRevB.92.104102} {\bibfield  {journal} {\bibinfo
  {journal} {Phys. Rev. B}\ }\textbf {\bibinfo {volume} {92}},\ \bibinfo
  {pages} {104102} (\bibinfo {year} {2015})}\BibitemShut {NoStop}%
\bibitem [{\citenamefont {Dieckmann}\ \emph {et~al.}(1983)\citenamefont
  {Dieckmann}, \citenamefont {Witt},\ and\ \citenamefont
  {Mason}}]{Dieckmann1983_V}%
  \BibitemOpen
  \bibfield  {author} {\bibinfo {author} {\bibfnamefont {R.}~\bibnamefont
  {Dieckmann}}, \bibinfo {author} {\bibfnamefont {C.~A.}\ \bibnamefont
  {Witt}},\ and\ \bibinfo {author} {\bibfnamefont {T.~O.}\ \bibnamefont
  {Mason}},\ }\href {https://doi.org/10.1002/bbpc.19830870609} {\bibfield
  {journal} {\bibinfo  {journal} {Berichte der Bunsengesellschaft für
  physikalische Chemie}\ }\textbf {\bibinfo {volume} {87}},\ \bibinfo {pages}
  {495} (\bibinfo {year} {1983})}\BibitemShut {NoStop}%
\bibitem [{\citenamefont {Dieckmann}\ and\ \citenamefont
  {Schmalzried}(1977)}]{Dieckmann1977_I}%
  \BibitemOpen
  \bibfield  {author} {\bibinfo {author} {\bibfnamefont {R.}~\bibnamefont
  {Dieckmann}}\ and\ \bibinfo {author} {\bibfnamefont {H.}~\bibnamefont
  {Schmalzried}},\ }\href {https://doi.org/10.1002/bbpc.19770810320} {\bibfield
   {journal} {\bibinfo  {journal} {Berichte der Bunsengesellschaft für
  physikalische Chemie}\ }\textbf {\bibinfo {volume} {81}},\ \bibinfo {pages}
  {344} (\bibinfo {year} {1977})}\BibitemShut {NoStop}%
\bibitem [{\citenamefont {Dieckmann}(1982)}]{Dieckmann1982_IV}%
  \BibitemOpen
  \bibfield  {author} {\bibinfo {author} {\bibfnamefont {R.}~\bibnamefont
  {Dieckmann}},\ }\href {https://doi.org/10.1002/bbpc.19820860205} {\bibfield
  {journal} {\bibinfo  {journal} {Berichte der Bunsengesellschaft für
  physikalische Chemie}\ }\textbf {\bibinfo {volume} {86}},\ \bibinfo {pages}
  {112} (\bibinfo {year} {1982})}\BibitemShut {NoStop}%
\bibitem [{\citenamefont {Banerjee}\ \emph {et~al.}(2021)\citenamefont
  {Banerjee}, \citenamefont {Kohnert}, \citenamefont {Holby},\ and\
  \citenamefont {Uberuaga}}]{Banerjee2021}%
  \BibitemOpen
  \bibfield  {author} {\bibinfo {author} {\bibfnamefont {A.}~\bibnamefont
  {Banerjee}}, \bibinfo {author} {\bibfnamefont {A.~A.}\ \bibnamefont
  {Kohnert}}, \bibinfo {author} {\bibfnamefont {E.~F.}\ \bibnamefont {Holby}},\
  and\ \bibinfo {author} {\bibfnamefont {B.~P.}\ \bibnamefont {Uberuaga}},\
  }\href {https://doi.org/10.1103/PhysRevMaterials.5.034410} {\bibfield
  {journal} {\bibinfo  {journal} {Phys. Rev. Mat.}\ }\textbf {\bibinfo {volume}
  {5}},\ \bibinfo {pages} {034410} (\bibinfo {year} {2021})}\BibitemShut
  {NoStop}%
\bibitem [{\citenamefont {Ma}\ \emph {et~al.}(2022)\citenamefont {Ma},
  \citenamefont {Souza~Filho}, \citenamefont {Bai}, \citenamefont {Schenk},
  \citenamefont {Patisson}, \citenamefont {Beck}, \citenamefont {van Bokhoven},
  \citenamefont {Willinger}, \citenamefont {Li}, \citenamefont {Xie},
  \citenamefont {Ponge}, \citenamefont {Zaefferer}, \citenamefont {Gault},
  \citenamefont {Mianroodi},\ and\ \citenamefont {Raabe}}]{Ma2022}%
  \BibitemOpen
  \bibfield  {author} {\bibinfo {author} {\bibfnamefont {Y.}~\bibnamefont
  {Ma}}, \bibinfo {author} {\bibfnamefont {I.~R.}\ \bibnamefont {Souza~Filho}},
  \bibinfo {author} {\bibfnamefont {Y.}~\bibnamefont {Bai}}, \bibinfo {author}
  {\bibfnamefont {J.}~\bibnamefont {Schenk}}, \bibinfo {author} {\bibfnamefont
  {F.}~\bibnamefont {Patisson}}, \bibinfo {author} {\bibfnamefont
  {A.}~\bibnamefont {Beck}}, \bibinfo {author} {\bibfnamefont {J.~A.}\
  \bibnamefont {van Bokhoven}}, \bibinfo {author} {\bibfnamefont {M.~G.}\
  \bibnamefont {Willinger}}, \bibinfo {author} {\bibfnamefont {K.}~\bibnamefont
  {Li}}, \bibinfo {author} {\bibfnamefont {D.}~\bibnamefont {Xie}}, \bibinfo
  {author} {\bibfnamefont {D.}~\bibnamefont {Ponge}}, \bibinfo {author}
  {\bibfnamefont {S.}~\bibnamefont {Zaefferer}}, \bibinfo {author}
  {\bibfnamefont {B.}~\bibnamefont {Gault}}, \bibinfo {author} {\bibfnamefont
  {J.~R.}\ \bibnamefont {Mianroodi}},\ and\ \bibinfo {author} {\bibfnamefont
  {D.}~\bibnamefont {Raabe}},\ }\href
  {https://doi.org/10.1016/j.scriptamat.2022.114571} {\bibfield  {journal}
  {\bibinfo  {journal} {Scripta Mat.}\ }\textbf {\bibinfo {volume} {213}},\
  \bibinfo {pages} {114571} (\bibinfo {year} {2022})}\BibitemShut {NoStop}%
\bibitem [{\citenamefont {Levy}\ \emph {et~al.}(2012)\citenamefont {Levy},
  \citenamefont {Giustetto},\ and\ \citenamefont {Hoser}}]{Levy2012}%
  \BibitemOpen
  \bibfield  {author} {\bibinfo {author} {\bibfnamefont {D.}~\bibnamefont
  {Levy}}, \bibinfo {author} {\bibfnamefont {R.}~\bibnamefont {Giustetto}},\
  and\ \bibinfo {author} {\bibfnamefont {A.}~\bibnamefont {Hoser}},\ }\href
  {https://doi.org/10.1007/s00269-011-0472-x} {\bibfield  {journal} {\bibinfo
  {journal} {Phys. Chem. Minerals}\ }\textbf {\bibinfo {volume} {39}},\
  \bibinfo {pages} {169} (\bibinfo {year} {2012})}\BibitemShut {NoStop}%
\bibitem [{\citenamefont {Arkharov}\ \emph {et~al.}(1972)\citenamefont
  {Arkharov}, \citenamefont {Bogoslovskii},\ and\ \citenamefont
  {Kuznetsov}}]{Arkharov1972}%
  \BibitemOpen
  \bibfield  {author} {\bibinfo {author} {\bibfnamefont {V.~I.}\ \bibnamefont
  {Arkharov}}, \bibinfo {author} {\bibfnamefont {V.~N.}\ \bibnamefont
  {Bogoslovskii}},\ and\ \bibinfo {author} {\bibfnamefont {E.~N.}\ \bibnamefont
  {Kuznetsov}},\ }\href@noop {} {\bibfield  {journal} {\bibinfo  {journal}
  {Neorganicheskie Mater.}\ }\textbf {\bibinfo {volume} {8}},\ \bibinfo {pages}
  {1982} (\bibinfo {year} {1972})}\BibitemShut {NoStop}%
\bibitem [{\citenamefont {Saito}(1965)}]{Saito1965}%
  \BibitemOpen
  \bibfield  {author} {\bibinfo {author} {\bibfnamefont {T.}~\bibnamefont
  {Saito}},\ }\href {https://doi.org/10.1246/bcsj.38.2008} {\bibfield
  {journal} {\bibinfo  {journal} {Bulletin of the Chemical Society of Japan}\
  }\textbf {\bibinfo {volume} {38}},\ \bibinfo {pages} {2008} (\bibinfo {year}
  {1965})}\BibitemShut {NoStop}%
\bibitem [{\citenamefont {Stukowski}(2009)}]{Stukowski2009}%
  \BibitemOpen
  \bibfield  {author} {\bibinfo {author} {\bibfnamefont {A.}~\bibnamefont
  {Stukowski}},\ }\href {https://doi.org/10.1088/0965-0393/18/1/015012}
  {\bibfield  {journal} {\bibinfo  {journal} {Modelling and Simulation in
  Materials Science and Engineering}\ }\textbf {\bibinfo {volume} {18}},\
  \bibinfo {pages} {015012} (\bibinfo {year} {2009})}\BibitemShut {NoStop}%
\bibitem [{\citenamefont {Lysogorskiy}\ \emph {et~al.}(2023)\citenamefont
  {Lysogorskiy}, \citenamefont {Bochkarev}, \citenamefont {Mrovec},\ and\
  \citenamefont {Drautz}}]{Lysogorskiy2023}%
  \BibitemOpen
  \bibfield  {author} {\bibinfo {author} {\bibfnamefont {Y.}~\bibnamefont
  {Lysogorskiy}}, \bibinfo {author} {\bibfnamefont {A.}~\bibnamefont
  {Bochkarev}}, \bibinfo {author} {\bibfnamefont {M.}~\bibnamefont {Mrovec}},\
  and\ \bibinfo {author} {\bibfnamefont {R.}~\bibnamefont {Drautz}},\ }\href
  {https://doi.org/10.1103/PhysRevMaterials.7.043801} {\bibfield  {journal}
  {\bibinfo  {journal} {Phys. Rev. Mater.}\ }\textbf {\bibinfo {volume} {7}},\
  \bibinfo {pages} {043801} (\bibinfo {year} {2023})}\BibitemShut {NoStop}%
\bibitem [{\citenamefont {Eder}\ \emph {et~al.}(2001)\citenamefont {Eder},
  \citenamefont {Terakura},\ and\ \citenamefont {Hafner}}]{Eder2001}%
  \BibitemOpen
  \bibfield  {author} {\bibinfo {author} {\bibfnamefont {M.}~\bibnamefont
  {Eder}}, \bibinfo {author} {\bibfnamefont {K.}~\bibnamefont {Terakura}},\
  and\ \bibinfo {author} {\bibfnamefont {J.}~\bibnamefont {Hafner}},\ }\href
  {https://doi.org/10.1103/PhysRevB.64.115426} {\bibfield  {journal} {\bibinfo
  {journal} {Phys. Rev. B}\ }\textbf {\bibinfo {volume} {64}},\ \bibinfo
  {pages} {115426} (\bibinfo {year} {2001})}\BibitemShut {NoStop}%
\bibitem [{\citenamefont {Dragoni}\ \emph {et~al.}(2018)\citenamefont
  {Dragoni}, \citenamefont {Daff}, \citenamefont {Csányi},\ and\ \citenamefont
  {Marzari}}]{Dragoni2018}%
  \BibitemOpen
  \bibfield  {author} {\bibinfo {author} {\bibfnamefont {D.}~\bibnamefont
  {Dragoni}}, \bibinfo {author} {\bibfnamefont {T.~D.}\ \bibnamefont {Daff}},
  \bibinfo {author} {\bibfnamefont {G.}~\bibnamefont {Csányi}},\ and\ \bibinfo
  {author} {\bibfnamefont {N.}~\bibnamefont {Marzari}},\ }\href
  {https://doi.org/10.1103/PhysRevMaterials.2.013808} {\bibfield  {journal}
  {\bibinfo  {journal} {Phys. Rev. Mat.}\ }\textbf {\bibinfo {volume} {2}},\
  \bibinfo {pages} {013808} (\bibinfo {year} {2018})}\BibitemShut {NoStop}%
\bibitem [{\citenamefont {Błoński}\ \emph {et~al.}(2005)\citenamefont
  {Błoński}, \citenamefont {Kiejna},\ and\ \citenamefont
  {Hafner}}]{Blonski2005}%
  \BibitemOpen
  \bibfield  {author} {\bibinfo {author} {\bibfnamefont {P.}~\bibnamefont
  {Błoński}}, \bibinfo {author} {\bibfnamefont {A.}~\bibnamefont {Kiejna}},\
  and\ \bibinfo {author} {\bibfnamefont {J.}~\bibnamefont {Hafner}},\ }\href
  {https://doi.org/10.1016/j.susc.2005.06.011} {\bibfield  {journal} {\bibinfo
  {journal} {Surf. Sci.}\ }\textbf {\bibinfo {volume} {590}},\ \bibinfo {pages}
  {88} (\bibinfo {year} {2005})}\BibitemShut {NoStop}%
\bibitem [{\citenamefont {Zhang}\ \emph {et~al.}(2023)\citenamefont {Zhang},
  \citenamefont {Jin}, \citenamefont {Han}, \citenamefont {Guo}, \citenamefont
  {Zhou}, \citenamefont {Liu},\ and\ \citenamefont {Shen}}]{Zhang2023}%
  \BibitemOpen
  \bibfield  {author} {\bibinfo {author} {\bibfnamefont {X.}~\bibnamefont
  {Zhang}}, \bibinfo {author} {\bibfnamefont {C.}~\bibnamefont {Jin}}, \bibinfo
  {author} {\bibfnamefont {S.}~\bibnamefont {Han}}, \bibinfo {author}
  {\bibfnamefont {P.}~\bibnamefont {Guo}}, \bibinfo {author} {\bibfnamefont
  {Y.}~\bibnamefont {Zhou}}, \bibinfo {author} {\bibfnamefont {W.}~\bibnamefont
  {Liu}},\ and\ \bibinfo {author} {\bibfnamefont {W.}~\bibnamefont {Shen}},\
  }\href {https://doi.org/10.1021/acs.inorgchem.3c01653} {\bibfield  {journal}
  {\bibinfo  {journal} {Inorganic Chem.}\ }\textbf {\bibinfo {volume} {62}},\
  \bibinfo {pages} {12111} (\bibinfo {year} {2023})}\BibitemShut {NoStop}%
\bibitem [{\citenamefont {Davenport}\ \emph {et~al.}(2000)\citenamefont
  {Davenport}, \citenamefont {Oblonsky}, \citenamefont {Ryan},\ and\
  \citenamefont {Toney}}]{Davenport2000}%
  \BibitemOpen
  \bibfield  {author} {\bibinfo {author} {\bibfnamefont {A.~J.}\ \bibnamefont
  {Davenport}}, \bibinfo {author} {\bibfnamefont {L.~J.}\ \bibnamefont
  {Oblonsky}}, \bibinfo {author} {\bibfnamefont {M.~P.}\ \bibnamefont {Ryan}},\
  and\ \bibinfo {author} {\bibfnamefont {M.~F.}\ \bibnamefont {Toney}},\ }\href
  {https://doi.org/10.1149/1.1393502} {\bibfield  {journal} {\bibinfo
  {journal} {J Electrochem. Soc.}\ }\textbf {\bibinfo {volume} {147}},\
  \bibinfo {pages} {2162} (\bibinfo {year} {2000})}\BibitemShut {NoStop}%
\bibitem [{\citenamefont {Zhou}\ \emph {et~al.}(2024)\citenamefont {Zhou},
  \citenamefont {Bienvenu}, \citenamefont {Wu}, \citenamefont {Kwiatkowski~da
  Silva}, \citenamefont {Ophus},\ and\ \citenamefont
  {Raabe}}]{Zhou2024_preprint}%
  \BibitemOpen
  \bibfield  {author} {\bibinfo {author} {\bibfnamefont {X.}~\bibnamefont
  {Zhou}}, \bibinfo {author} {\bibfnamefont {B.}~\bibnamefont {Bienvenu}},
  \bibinfo {author} {\bibfnamefont {Y.}~\bibnamefont {Wu}}, \bibinfo {author}
  {\bibfnamefont {A.}~\bibnamefont {Kwiatkowski~da Silva}}, \bibinfo {author}
  {\bibfnamefont {C.}~\bibnamefont {Ophus}},\ and\ \bibinfo {author}
  {\bibfnamefont {D.}~\bibnamefont {Raabe}},\ }\href@noop {} {\bibfield
  {journal} {\bibinfo  {journal} {Preprint}\ } (\bibinfo {year}
  {2024})}\BibitemShut {NoStop}%
\bibitem [{\citenamefont {Kresse}\ and\ \citenamefont
  {Furthmüller}(1996)}]{Kresse1996}%
  \BibitemOpen
  \bibfield  {author} {\bibinfo {author} {\bibfnamefont {G.}~\bibnamefont
  {Kresse}}\ and\ \bibinfo {author} {\bibfnamefont {J.}~\bibnamefont
  {Furthmüller}},\ }\href {https://doi.org/10.1016/0927-0256(96)00008-0}
  {\bibfield  {journal} {\bibinfo  {journal} {Comput. Mater. Sci.}\ }\textbf
  {\bibinfo {volume} {6}},\ \bibinfo {pages} {15} (\bibinfo {year}
  {1996})}\BibitemShut {NoStop}%
\bibitem [{\citenamefont {Blöchl}(1994)}]{Blochl1994}%
  \BibitemOpen
  \bibfield  {author} {\bibinfo {author} {\bibfnamefont {P.~E.}\ \bibnamefont
  {Blöchl}},\ }\href {https://doi.org/10.1103/PhysRevB.50.17953} {\bibfield
  {journal} {\bibinfo  {journal} {Phys. Rev. B}\ }\textbf {\bibinfo {volume}
  {50}},\ \bibinfo {pages} {17953} (\bibinfo {year} {1994})}\BibitemShut
  {NoStop}%
\bibitem [{\citenamefont {Thompson}\ \emph {et~al.}(2022)\citenamefont
  {Thompson}, \citenamefont {Aktulga}, \citenamefont {Berger}, \citenamefont
  {Bolintineanu}, \citenamefont {Brown}, \citenamefont {Crozier}, \citenamefont
  {in~'t Veld}, \citenamefont {Kohlmeyer}, \citenamefont {Moore}, \citenamefont
  {Nguyen}, \citenamefont {Shan}, \citenamefont {Stevens}, \citenamefont
  {Tranchida}, \citenamefont {Trott},\ and\ \citenamefont
  {Plimpton}}]{Thompson2022}%
  \BibitemOpen
  \bibfield  {author} {\bibinfo {author} {\bibfnamefont {A.~P.}\ \bibnamefont
  {Thompson}}, \bibinfo {author} {\bibfnamefont {H.~M.}\ \bibnamefont
  {Aktulga}}, \bibinfo {author} {\bibfnamefont {R.}~\bibnamefont {Berger}},
  \bibinfo {author} {\bibfnamefont {D.~S.}\ \bibnamefont {Bolintineanu}},
  \bibinfo {author} {\bibfnamefont {W.~M.}\ \bibnamefont {Brown}}, \bibinfo
  {author} {\bibfnamefont {P.~S.}\ \bibnamefont {Crozier}}, \bibinfo {author}
  {\bibfnamefont {P.~J.}\ \bibnamefont {in~'t Veld}}, \bibinfo {author}
  {\bibfnamefont {A.}~\bibnamefont {Kohlmeyer}}, \bibinfo {author}
  {\bibfnamefont {S.~G.}\ \bibnamefont {Moore}}, \bibinfo {author}
  {\bibfnamefont {T.~D.}\ \bibnamefont {Nguyen}}, \bibinfo {author}
  {\bibfnamefont {R.}~\bibnamefont {Shan}}, \bibinfo {author} {\bibfnamefont
  {M.~J.}\ \bibnamefont {Stevens}}, \bibinfo {author} {\bibfnamefont
  {J.}~\bibnamefont {Tranchida}}, \bibinfo {author} {\bibfnamefont
  {C.}~\bibnamefont {Trott}},\ and\ \bibinfo {author} {\bibfnamefont {S.~J.}\
  \bibnamefont {Plimpton}},\ }\href {https://doi.org/10.1016/j.cpc.2021.108171}
  {\bibfield  {journal} {\bibinfo  {journal} {Comput. Phys. Commun.}\ }\textbf
  {\bibinfo {volume} {271}},\ \bibinfo {pages} {108171} (\bibinfo {year}
  {2022})}\BibitemShut {NoStop}%
\bibitem [{\citenamefont {Sadigh}\ \emph {et~al.}(2012)\citenamefont {Sadigh},
  \citenamefont {Erhart}, \citenamefont {Stukowski}, \citenamefont {Caro},
  \citenamefont {Martinez},\ and\ \citenamefont {Zepeda-Ruiz}}]{Sadigh2012}%
  \BibitemOpen
  \bibfield  {author} {\bibinfo {author} {\bibfnamefont {B.}~\bibnamefont
  {Sadigh}}, \bibinfo {author} {\bibfnamefont {P.}~\bibnamefont {Erhart}},
  \bibinfo {author} {\bibfnamefont {A.}~\bibnamefont {Stukowski}}, \bibinfo
  {author} {\bibfnamefont {A.}~\bibnamefont {Caro}}, \bibinfo {author}
  {\bibfnamefont {E.}~\bibnamefont {Martinez}},\ and\ \bibinfo {author}
  {\bibfnamefont {L.}~\bibnamefont {Zepeda-Ruiz}},\ }\href
  {https://doi.org/10.1103/PhysRevB.85.184203} {\bibfield  {journal} {\bibinfo
  {journal} {Phys. Rev. B}\ }\textbf {\bibinfo {volume} {85}},\ \bibinfo
  {pages} {184203} (\bibinfo {year} {2012})}\BibitemShut {NoStop}%
\bibitem [{\citenamefont {Adams}\ \emph {et~al.}(2006)\citenamefont {Adams},
  \citenamefont {Agosta}, \citenamefont {Leisure},\ and\ \citenamefont
  {Ledbetter}}]{Adams2006}%
  \BibitemOpen
  \bibfield  {author} {\bibinfo {author} {\bibfnamefont {J.~J.}\ \bibnamefont
  {Adams}}, \bibinfo {author} {\bibfnamefont {D.~S.}\ \bibnamefont {Agosta}},
  \bibinfo {author} {\bibfnamefont {R.~G.}\ \bibnamefont {Leisure}},\ and\
  \bibinfo {author} {\bibfnamefont {H.}~\bibnamefont {Ledbetter}},\ }\href
  {https://doi.org/10.1063/1.2365714} {\bibfield  {journal} {\bibinfo
  {journal} {J. Applied Phys.}\ }\textbf {\bibinfo {volume} {100}},\ \bibinfo
  {pages} {113530} (\bibinfo {year} {2006})}\BibitemShut {NoStop}%
\bibitem [{\citenamefont {Wollenberger}(1996)}]{Wollenberger1996}%
  \BibitemOpen
  \bibfield  {author} {\bibinfo {author} {\bibfnamefont {H.~J.}\ \bibnamefont
  {Wollenberger}},\ }\href@noop {} {\bibfield  {journal} {\bibinfo  {journal}
  {Physical Metall.}\ }\textbf {\bibinfo {volume} {2}} (\bibinfo {year}
  {1996})}\BibitemShut {NoStop}%
\bibitem [{\citenamefont {Nazarov}\ \emph {et~al.}(2010)\citenamefont
  {Nazarov}, \citenamefont {Hickel},\ and\ \citenamefont
  {Neugebauer}}]{Nazarov2010}%
  \BibitemOpen
  \bibfield  {author} {\bibinfo {author} {\bibfnamefont {R.}~\bibnamefont
  {Nazarov}}, \bibinfo {author} {\bibfnamefont {T.}~\bibnamefont {Hickel}},\
  and\ \bibinfo {author} {\bibfnamefont {J.}~\bibnamefont {Neugebauer}},\
  }\href {https://doi.org/10.1103/PhysRevB.82.224104} {\bibfield  {journal}
  {\bibinfo  {journal} {Physical Review B}\ }\textbf {\bibinfo {volume} {82}},\
  \bibinfo {pages} {224104} (\bibinfo {year} {2010})}\BibitemShut {NoStop}%
\bibitem [{\citenamefont {Ventelon}\ and\ \citenamefont
  {Willaime}(2010)}]{Ventelon2010}%
  \BibitemOpen
  \bibfield  {author} {\bibinfo {author} {\bibfnamefont {L.}~\bibnamefont
  {Ventelon}}\ and\ \bibinfo {author} {\bibfnamefont {F.}~\bibnamefont
  {Willaime}},\ }\href {https://doi.org/10.1080/14786431003668793} {\bibfield
  {journal} {\bibinfo  {journal} {Philos. Mag.}\ }\textbf {\bibinfo {volume}
  {90}},\ \bibinfo {pages} {1063} (\bibinfo {year} {2010})}\BibitemShut
  {NoStop}%
\bibitem [{\citenamefont {Bienvenu}\ \emph {et~al.}(2022)\citenamefont
  {Bienvenu}, \citenamefont {Dezerald}, \citenamefont {Rodney},\ and\
  \citenamefont {Clouet}}]{Bienvenu2022}%
  \BibitemOpen
  \bibfield  {author} {\bibinfo {author} {\bibfnamefont {B.}~\bibnamefont
  {Bienvenu}}, \bibinfo {author} {\bibfnamefont {L.}~\bibnamefont {Dezerald}},
  \bibinfo {author} {\bibfnamefont {D.}~\bibnamefont {Rodney}},\ and\ \bibinfo
  {author} {\bibfnamefont {E.}~\bibnamefont {Clouet}},\ }\href
  {https://doi.org/10.1016/j.actamat.2022.118098} {\bibfield  {journal}
  {\bibinfo  {journal} {Acta Mater.}\ }\textbf {\bibinfo {volume} {236}},\
  \bibinfo {pages} {118098} (\bibinfo {year} {2022})}\BibitemShut {NoStop}%
\bibitem [{\citenamefont {Bleskov}\ \emph {et~al.}(2016)\citenamefont
  {Bleskov}, \citenamefont {Hickel}, \citenamefont {Neugebauer},\ and\
  \citenamefont {Ruban}}]{Bleskov2016}%
  \BibitemOpen
  \bibfield  {author} {\bibinfo {author} {\bibfnamefont {I.}~\bibnamefont
  {Bleskov}}, \bibinfo {author} {\bibfnamefont {T.}~\bibnamefont {Hickel}},
  \bibinfo {author} {\bibfnamefont {J.}~\bibnamefont {Neugebauer}},\ and\
  \bibinfo {author} {\bibfnamefont {A.}~\bibnamefont {Ruban}},\ }\href
  {https://doi.org/10.1103/PhysRevB.93.214115} {\bibfield  {journal} {\bibinfo
  {journal} {Phys. Rev. B}\ }\textbf {\bibinfo {volume} {93}},\ \bibinfo
  {pages} {214115} (\bibinfo {year} {2016})}\BibitemShut {NoStop}%
\end{thebibliography}%

\clearpage

\appendix
\noindent
\textbf{Supplementary Information}
\noindent

\makeatletter
\@addtoreset{equation}{section}
\@addtoreset{table}{section}
\makeatother

\section{Additional validation on pure Fe}
\label{app:pure_Fe}

We present in this section extensive additional validation of the Fe-O ACE potential performed on pure Fe, including bulk and defect properties.

\subsection{Bulk properties}

Tables \ref{tab:prop_Fe_BCC}, \ref{tab:prop_Fe_FCC} and \ref{tab:prop_Fe_HCP} summarize elastic constants $C_{ij}$ of BCC, FCC and HCP Fe respectively. For each, we considered different magnetic orders, namely FM and NM for BCC, AFDL and NM for FCC, and AF and NM for HCP. When available or computed for this study, DFT references are included. Available experimental data are also presented. For the different crystal structures and magnetic orders considered here, we note a very good agreement between the ACE potential and reference data.

We report some problems for the FM magnetic order of FCC Fe caused by the two low-spin and high-spin states that the potential is trying to reproduce. Since the magnetic model used for parameterization of the potential does not allow to distinguish between these two states, this results in instability of the FM state of FCC Fe at zero pressure. However, this is not the magnetic ground-state of FCC Fe, which is the antiferromagnetic double layer (AFDL), whose equilibrium properties are also presented in Tab. \ref{tab:prop_Fe_FCC}.

\begin{figure*}[!htb]    
    \hspace{-14mm}
    \includegraphics[trim = 0mm 0mm 0mm 0mm, clip, width=0.75\linewidth]{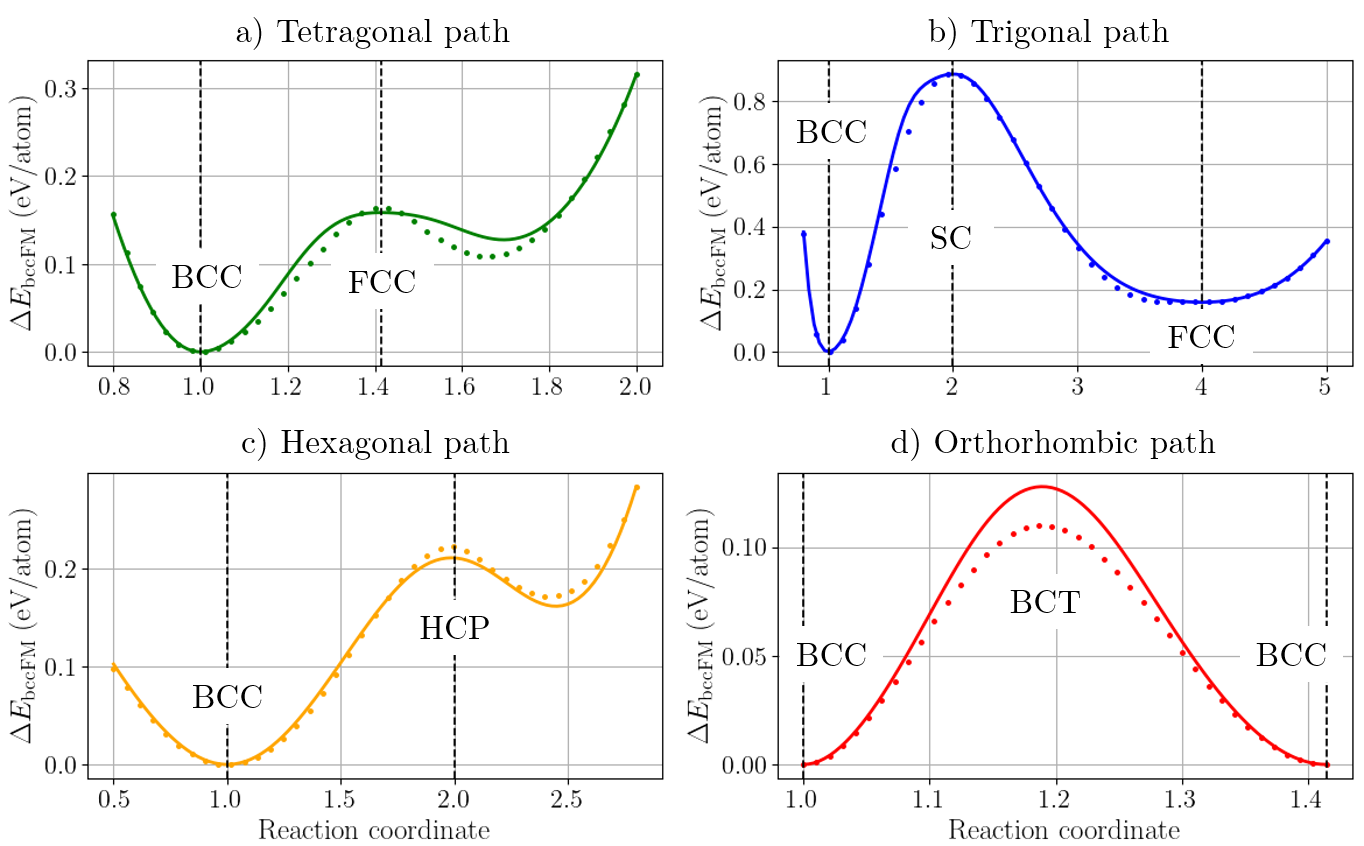}
    \caption{Transformation paths of Fe (FM magnetic order): a) tetragonal (or Bain) path from BCC to FCC ; b) trigonal path from BCC to simple cubic (SC) to FCC ; c) hexagonal path from BCC to HCP ; d) orthorhombic path from BCC to body-centered tetragonal (BCT).}
    \label{fig:paths_Fe}
\end{figure*}

We also present in Fig. \ref{fig:paths_Fe} transformation paths between different crystal structures for pure Fe with a FM order. Along these paths, the coordination, bond distances and angles change significantly, making them a sensitive test when assessing the robustness of any interatomic potential \cite{Lysogorskiy2021}. We again report a very good agreement between the DFT reference data and the results obtained with the ACE potential, demonstrating its robustness and accuracy when describing highly distorted atomic environments.

\begin{figure*}[!htb]    
    \hspace{-10mm}
    \includegraphics[trim = 0mm 0mm 0mm 0mm, clip, width=0.65\linewidth]{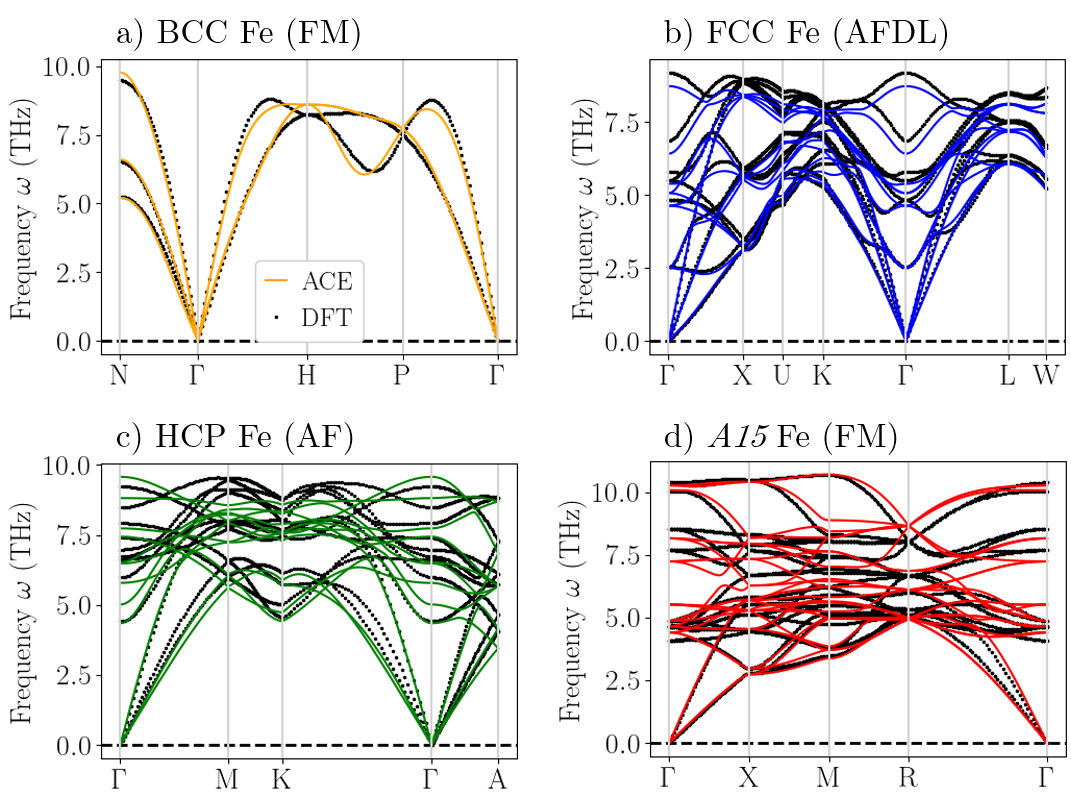}
    \caption{Phonon spectrum for a) BCC Fe (FM magnetic order), b) FCC Fe (AFDL magnetic order), c) HCP Fe (AF magnetic order), and d) $A15$ Fe (FM magnetic order). DFT data (black dots) are compared to the results of the ACE potential (solid color lines).}
    \label{fig:phonons_Fe}
\end{figure*}

Finally, phonon spectra for the four BCC FM, FCC AFDL, HCP AF and $A15$ FM phases of pure Fe are presented in Fig. \ref{fig:phonons_Fe} and compared to results obtained using DFT. We note a very good agreement between the two methods, demonstrating that the ACE potential is also able to capture vibrational properties in a wide range of atomic configurations in the case of pure Fe.

\begin{center}
    \begin{table}[!htb]
        \caption{Bulk properties (lattice parameter $a_0$, bulk modulus $B_0$ and elastic constants $C_{ij}$) of the FM and NM magnetic orders of BCC Fe, and defect formation energies (vacancy $E_{f}^{\text{vac.}}$, self-interstitial atom with various dumbbell configurations $E_{f}^{\text{dhkl}}$, and surface energies $\gamma^{\text{\hkl{hkl}}}$, where $hkl$ denote the Miller indices of the crystallographic orientation considered) in the FM ground-state only.}
        \label{tab:prop_Fe_BCC}
        \centering
        \begin{tabular}{l c c l}
              & ACE & DFT \,\, & Expt. \\
            \hline
            \multicolumn{4}{l}{Fe, BCC FM} \\
            \hline
            $a_0$ ({\AA}) & 2.83 & 2.83 & 2.86 \cite{Basinski1955} \\
            $B_0$ (GPa) & 171 & 191 \cite{Rinaldi2024} & 170 \cite{Adams2006} \\
            $C_{11}$ (GPa) & 261 & 283 \cite{Rinaldi2024} & 240 \cite{Adams2006} \\
            $C_{12}$ (GPa) & 125 & 145 \cite{Rinaldi2024} & 136 \cite{Adams2006} \\
            $C_{44}$ (GPa) & 107 & 104 \cite{Rinaldi2024} & 121 \cite{Adams2006} \\
            $E_{f}^{\text{vac.}}$ (eV) & 2.1 & 2.2 & $2.0 \pm 0.2$ \cite{Schepper1983} \\
            $E_{f}^{\text{d110}}$ (eV) & 4.4 & 4.9 \cite{Rinaldi2024} & 4.7\,-\,5.0 \cite{Wollenberger1996} \\
            $E_{f}^{\text{d100}}$ (eV) & 5.3 & 5.3 \cite{Rinaldi2024} & / \\
            $E_{f}^{\text{d111}}$ (eV) & 4.4 & 4.9 \cite{Rinaldi2024} & / \\
            $\gamma^{\text{\hkl{100}}}$ (J/m$^2$) & 2.62 & 2.49 & / \\
            $\gamma^{\text{\hkl{110}}}$ (J/m$^2$) & 2.48 & 2.42 & / \\
            $\gamma^{\text{\hkl{112}}}$ (J/m$^2$) & 2.68 & 2.56 & / \\
            $\gamma^{\text{\hkl{111}}}$ (J/m$^2$) & 2.77 & 2.69 & / \\
            \hline
            \multicolumn{4}{l}{Fe, BCC NM} \\
            \hline
            $\Delta E$ (eV/atom) & 0.47 & 0.48 & / \\
            $a_0$ ({\AA}) & 2.75 & 2.75 & / \\
            $B_0$ (GPa) & 280 & 270 \cite{Rinaldi2024} & / \\
            $C_{11}$ (GPa) & 39 & 87 \cite{Rinaldi2024} & / \\
            $C_{12}$ (GPa) & 400 & 361 \cite{Rinaldi2024} & / \\
            $C_{44}$ (GPa) & 189 & 180 \cite{Rinaldi2024} & / \\
            \hline
        \end{tabular}
    \end{table}
\end{center}

\begin{center}
    \begin{table}[!htb]
        \caption{Same as Tab. \ref{tab:prop_Fe_BCC} for the FCC phase of Fe.}
        \label{tab:prop_Fe_FCC}
        \centering
        \begin{tabular}{l c c l}
              & ACE & DFT \,\, & Expt. \\
            \hline
            \multicolumn{4}{l}{Fe, FCC AFDL} \\
            \hline
            $\Delta E$ (eV/atom) & 0.11 & 0.11 & / \\
            $a_0$ ({\AA}) & 3.55 & 3.55 & / \\
            $B_0$ (GPa) & 143 &  & 146 \cite{Dorogokupets2017} \\
            $C_{11}$ (GPa) & 331 & / & / \\
            $C_{12}$ (GPa) & 33 & / & / \\
            $C_{44}$ (GPa) & 59 & / & / \\
            $E_{f}^{\text{vac.}}$ (eV) & 1.60 & 1.82 \cite{Nazarov2010} & / \\
            \hline
            \multicolumn{4}{l}{Fe, FCC NM} \\
            \hline
            $\Delta E$ (eV/atom) & 0.16 & 0.17 & / \\
            $a_0$ ({\AA}) & 3.45 & 3.45 \cite{Rinaldi2024} & / \\
            $B_0$ (GPa) & 288 & 281 \cite{Rinaldi2024} & 146 \cite{Dorogokupets2017} \\
            $C_{11}$ (GPa) & 465 & 414 \cite{Rinaldi2024} & / \\
            $C_{12}$ (GPa) & 199 & 214 \cite{Rinaldi2024} & / \\
            $C_{44}$ (GPa) & 201 & 240 \cite{Rinaldi2024} & / \\
            $E_{f}^{\text{vac.}}$ (eV) & 2.27 & 2.27 \cite{Nazarov2010} & / \\
            \hline
        \end{tabular}
    \end{table}
\end{center}

\begin{center}
    \begin{table}[!htb]
        \caption{Same as Tab. \ref{tab:prop_Fe_BCC} for the HCP phase of Fe.}
        \label{tab:prop_Fe_HCP}
        \centering
        \begin{tabular}{l c c l}
              & ACE & DFT \,\, & Expt. \\
            \hline
            \multicolumn{4}{l}{Fe, HCP AF} \\
            \hline
            $\Delta E$ (eV/atom) & 0.06 & 0.06 & / \\
            $V_0$ ({\AA}$^3$/atom) & 10.53 & 10.55 & / \\
            $B_0$ (GPa) & 262 & 199 & / \\
            $C_{11}$ (GPa) & 484 & / & / \\
            $C_{12}$ (GPa) & 124 & / & / \\
            $C_{44}$ (GPa) & 109 & / & / \\
            $E_{f}^{\text{vac.}}$ (eV) & 1.76 & 2.27 & / \\
            \hline
            \multicolumn{4}{l}{Fe, HCP NM} \\
            \hline
            $\Delta E$ (eV/atom) & 0.10 & 0.10 & / \\
            $V_0$ ({\AA}$^3$/atom) & 10.14 & 10.18 &  \\
            $B_0$ (GPa) & 313 & 291 & / \\
            $C_{11}$ (GPa) & 545 & / & / \\
            $C_{12}$ (GPa) & 157 & / & / \\
            $C_{44}$ (GPa) & 180 & / & / \\
            $E_{f}^{\text{vac.}}$ (eV) & 2.07 & 2.45 & / \\
            \hline
        \end{tabular}
    \end{table}
\end{center}

\subsection{Defect properties}

Defect properties are presented only in the stable structures, which are BCC FM, FCC AFDL and NM, HCP AFII and NM. The stability of the different crystal structures are assessed through the existence of imaginary phonon modes.

We present in Tables \ref{tab:prop_Fe_BCC}, \ref{tab:prop_Fe_FCC} and \ref{tab:prop_Fe_HCP} the vacancy formation energies for the different crystal structures of pure Fe. We note a very good agreement with available DFT references in the three phases of pure Fe, with a lower formation energy in the FCC and HCP phases than in the BCC phase. For the FM magnetic order of BCC Fe, we also compare in Tab. \ref{tab:prop_Fe_BCC} the surface energies of different crystal orientations predicted by ACE with DFT reference data, between which we note a very satisfactory agreement. In the same phase, we also computed the formation energy of self-interstitial Fe atoms having different dumbbell configurations (see Tab. \ref{tab:prop_Fe_BCC}). Values predicted by ACE are in very good agreement with previously reported DFT values, for instance reproducing the very close energies of the two \hkl<110> and \hkl<111> configurations.

We stress that some of the presented properties are usually difficult to accurately reproduce for interatomic potentials, for instance vacancy formation energies, self-interstitials and the hierarchy between different configurations, and the almost degeneracy between the two \hkl{100} and \hkl{110} surfaces of BCC FM.

\begin{figure*}[!htb]    
    \hspace{-5mm}
    \includegraphics[trim = 0mm 0mm 0mm 0mm, clip, width=0.95\linewidth]{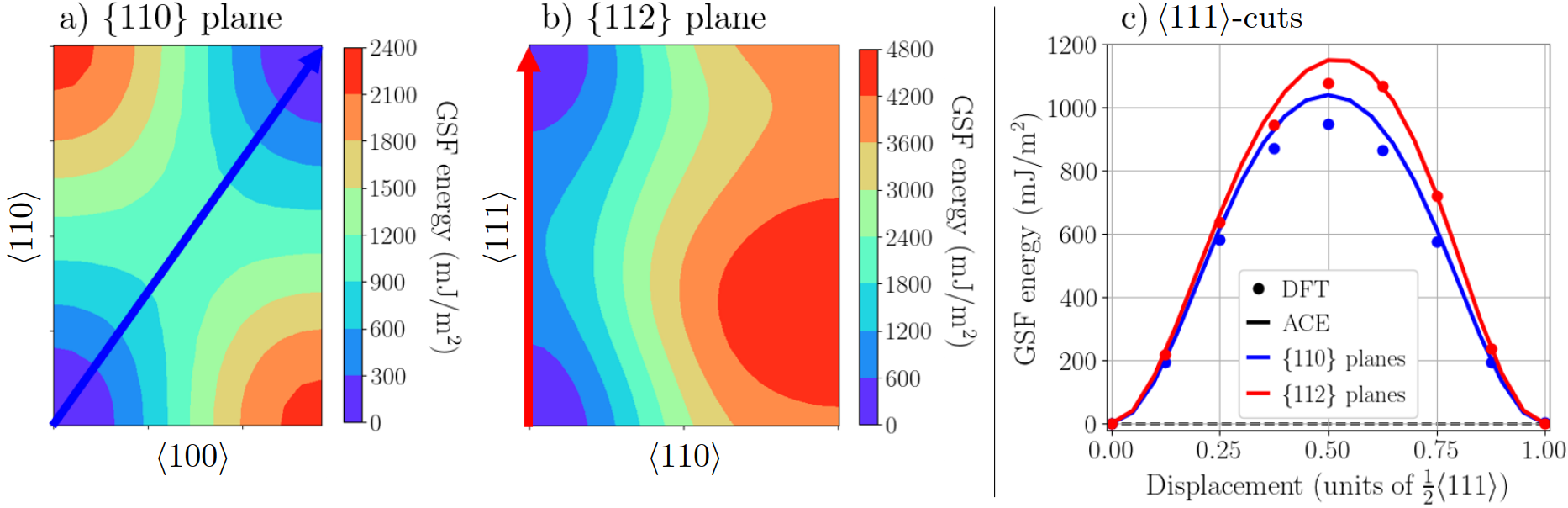}
    \caption{Generalized stacking faults in a a) \hkl{110} and b) \hkl{112} plane of BCC Fe (FM magnetic order) obtained using the $\text{ACE}_{\text{Fe-O}}$ potential. c) Cuts of the $\gamma$-surfaces along a $\sfrac{1}{2}\hkl<111>$ direction in \hkl{110} (blue) and \hkl{112} (red) planes.}
    \label{fig:GSF_bccFe}
\end{figure*}

We also further validate the present ACE potential on more complicated defects, namely dislocations. As a first step towards rationalizing the motion of dislocations in a crystal, the generalized stacking faults give interesting information on the ease to shear different crystallographic planes (assimilated to the glide planes of dislocations) by a given vector (assimilated to the Burgers vector of dislocations). These are presented in Fig. \ref{fig:GSF_bccFe} in the two \hkl{110} and \hkl{112} planes of BCC Fe, which are the two main glide planes of dislocations in BCC metals.

Cuts along the \hkl<111> directions contained in the two planes are also presented in Fig. \ref{fig:GSF_bccFe}\,c, where $\sfrac{1}{2}\hkl<111>$ represent the main Burgers vector of dislocations in BCC crystals. We note that compared to DFT data, the ACE potential captures very well the shearing of the lattice in the two considered planes, also reproducing the asymmetry in the \hkl{112} planes (red line in Fig. \ref{fig:GSF_bccFe}\,c).


\begin{figure*}[!htb]    
    \hspace{-14mm}
    \includegraphics[trim = 0mm 0mm 0mm 0mm, clip, width=0.75\linewidth]{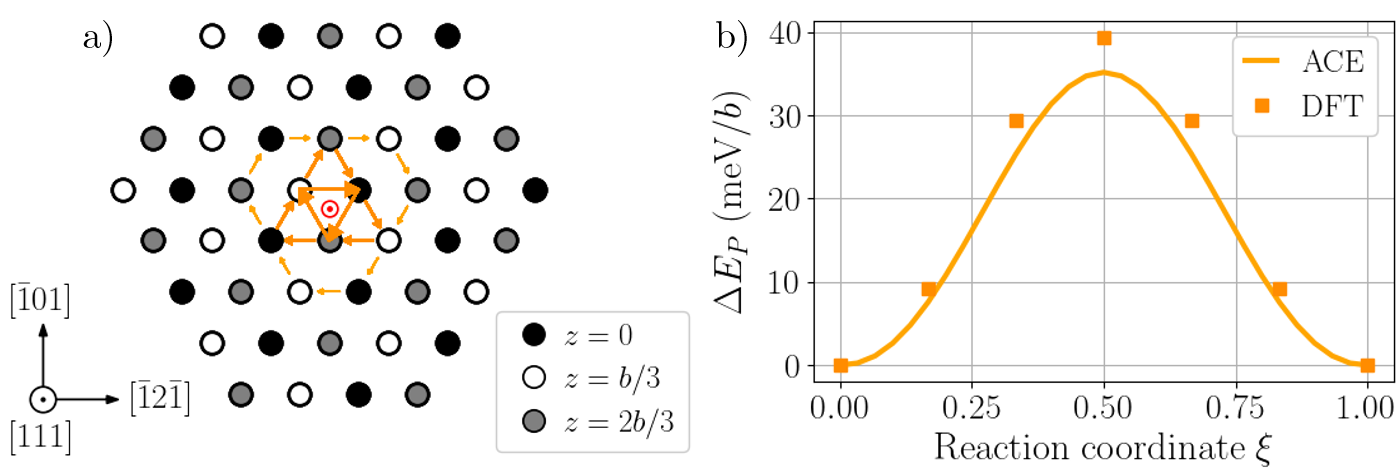}
    \caption{a) Core structure of a $\sfrac{1}{2}\hkl<111>$ screw dislocation predicted by the ACE potential plotted as a differential displacement map along the \hkl[111] direction. b) Peierls barrier opposing glide of a $\sfrac{1}{2}\hkl<111>$ in a \hkl{110} plane between two adjacent equilibrium positions. DFT data is taken from Ref. \cite{Bienvenu2022}.}
    \label{fig:b111_screw}
\end{figure*}

We present in Fig. \ref{fig:b111_screw} the core structure and the Peierls energy barrier for the $\sfrac{1}{2}\hkl<111>$ screw dislocation in BCC Fe (FM magnetic order). Plasticity of BCC metals is governed at low temperature by the motion of these $\sfrac{1}{2}\hkl<111>$ screw dislocations through the crystal, due to friction they experience with the lattice. Thus, in order to be able to accurately describe the plastic deformation of BCC Fe, it is important that the ACE potential is also able to describe its core structure and the energy barrier associated with its motion, namely the Peierls barrier. We note that the Fe-O ACE potential accurately predicts the compact core structure for the $\sfrac{1}{2}\hkl<111>$ screw dislocation, with a Peierls barrier of very similar height compared to DFT reference. We also note that these properties are usually hard to correctly reproduce by classical EAM interatomic potentials \cite{Ventelon2010}.

\begin{figure*}[!htb]    
    \hspace{-14mm}
    \includegraphics[trim = 0mm 0mm 0mm 0mm, clip, width=0.7\linewidth]{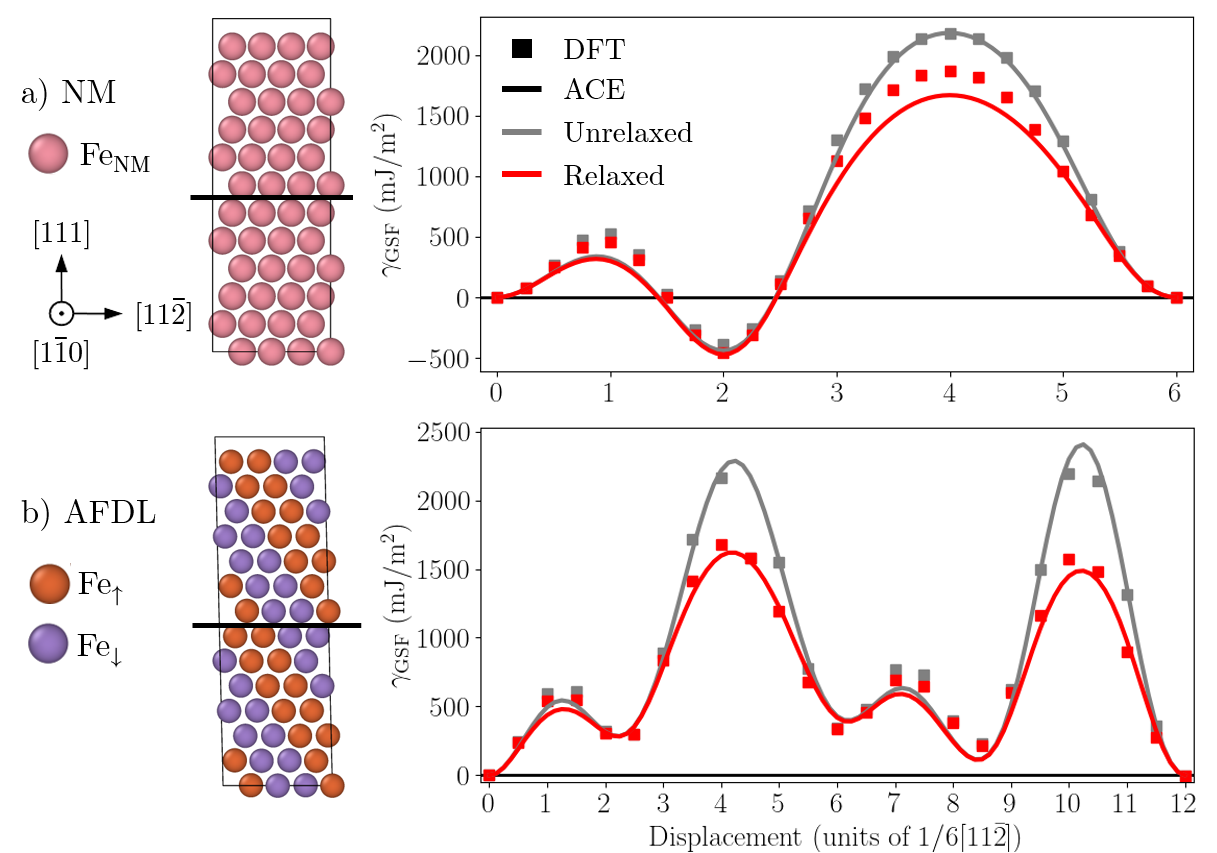}
    \caption{Cut of the \hkl{111} $\gamma$-surface of FCC Fe along a \hkl<112> direction in the a) NM and b) AFDL magnetic states. The path for the AFDL structure is twice longer than the NM case due to the breaking of the periodicity of the magnetic order by a \hkl<112> shear in a \hkl{111} plane.}
    \label{fig:SF_fccFe}
\end{figure*}

Finally, we present in Fig. \ref{fig:SF_fccFe} the generalized stacking fault of FCC Fe in the \hkl(111) plane, cut along the \hkl[11-2] direction. First of all, we note a very good agreement between the predictions of the ACE potential and DFT, also in agreement with a previous DFT study \cite{Bleskov2016}. Looking at the line cut, the stable stacking fault (located at a shearing of $2/6\,\hkl[11-2]$) corresponds to a local HCP structure near the fault plane. The negative stable stacking fault energy obtained in FCC Fe with a NM order indicates that the HCP structure is more stable than FCC. This is not the case in the AF magnetic order, where the stable stacking fault has a positive energy, directly linked to the dissociation of dislocations in FCC Fe.

\section{Additional validation on oxides}

\begin{figure*}[!htb]    
    \hspace{-10mm}
    \includegraphics[trim = 0mm 0mm 0mm 0mm, clip, width=0.65\linewidth]{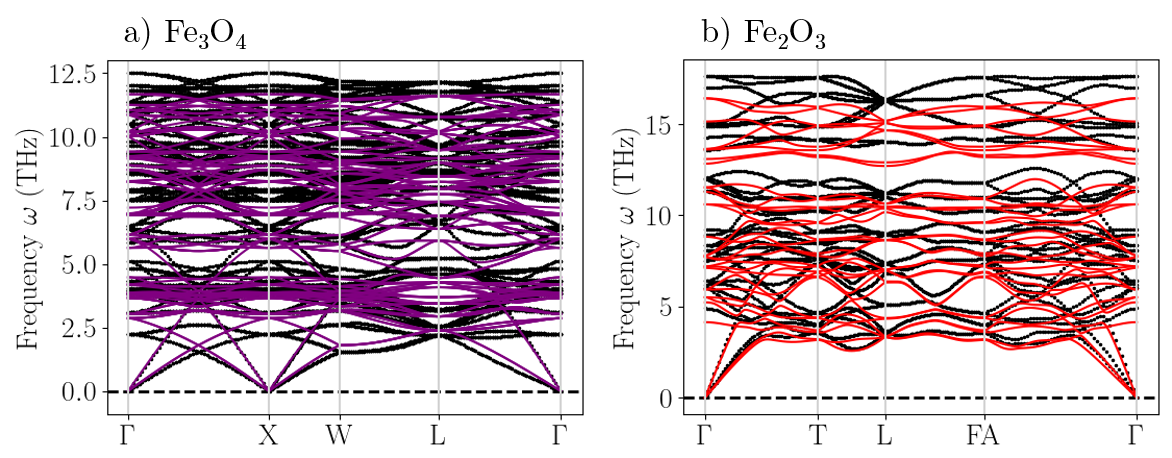}
    \caption{Phonon spectrum for a) {\magnetite} (ferrimagnetic order) and b) {\hematite} (AF magnetic order). DFT data (black dots) are compared to the results of the $\text{ACE}_{\text{Fe-O}}$ potential (solid color lines).}
    \label{fig:phonons_oxides}
\end{figure*}

As for iron oxides, we present in Fig. \ref{fig:phonons_oxides} the phonon spectrum of {\magnetite} and {\hematite}, comparing the results obtained using ACE and DFT. We note a very satisfactory agreement between the two methods for both oxides, showing the ability of the ACE potential to also describe vibrational properties of iron oxides.

\section{Effect of DFT xc-functional on bulk properties of pure Fe}

\begin{figure*}[!htb]    
    \hspace{-14mm}
    \includegraphics[trim = 0mm 0mm 0mm 0mm, clip, width=1\linewidth]{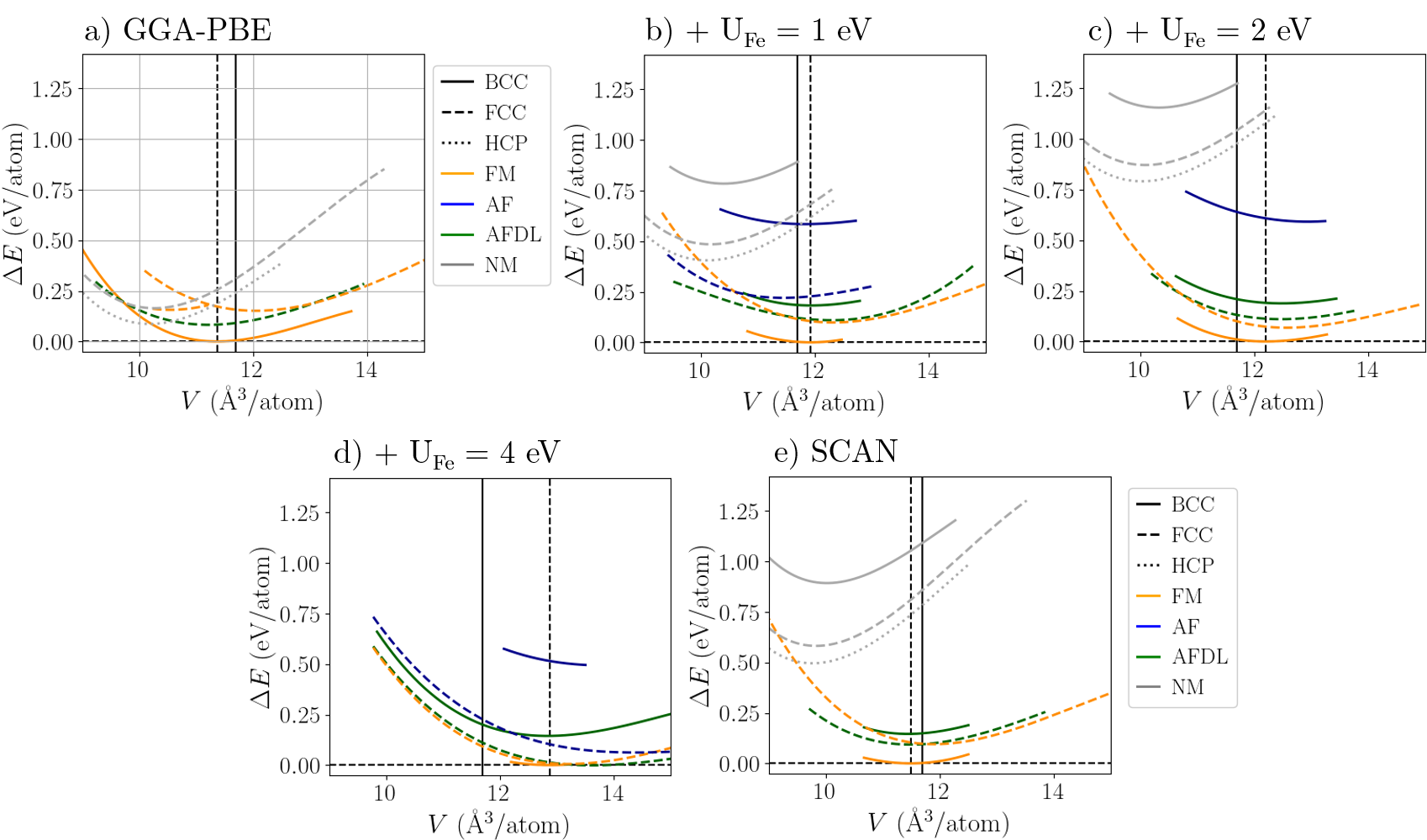}
    \caption{Energy as a function of atomic volume for different crystal structures and magnetic orders of pure Fe using different DFT functionals: a) standard GGA-PBE, b) $\text{GGA-PBE}+U_{\text{Fe}}=1\,$eV, c) $\text{GGA-PBE}+U_{\text{Fe}}=2\,$eV, d) $\text{GGA-PBE}+U_{\text{Fe}}=4\,$eV, and e) SCAN.}
    \label{fig:xc_Fe}
\end{figure*}

As mentioned in the main text, we investigated the effect of using different DFT xc-functionals on the bulk properties of pure Fe. The results are presented in Fig. \ref{fig:xc_Fe}, considering GGA-PBE, different values for the $U_{\text{Fe}}$ correction added on the $3d$ orbitals of Fe atoms, and the SCAN meta-GGA xc-functionals.

Comparing pure GGA-PBE results with {\dftu}, we observe that increasing the values of the $U_{\text{Fe}}$ correction gradually decreases the energy difference between the BCC FM and the FCC AFDL phases of pure Fe until the two phases have almost the same energy for $U_{\text{Fe}}=4$\,eV (see Fig. \ref{fig:xc_Fe}\,d). Increasing $U_{\text{Fe}}$ also raises the energy of all NM phases of Fe, regardless of the crystal structure. On the other side, the use of the SCAN xc-functional yields results in a rather good agreement with GGA-PBE for the two BCC and FCC phases. However, the energy of the HCP NM phase is predicted rather high, pushing the pressure at which BCC Fe would transform to HCP.

\end{document}